\documentclass[nofootinbib,aps,prd,10pt,superscriptaddress,showpacs,preprintnumbers,eqsecnum]{revtex4-1}

\usepackage{amsmath}
\usepackage{multirow}
\usepackage{graphicx}
\usepackage{mathrsfs}

\usepackage{amsfonts}
\usepackage{amssymb}
\usepackage[dvips]{color}

\begin{document}
\newcommand\be{\begin{equation}}
\newcommand\ee{\end{equation}}
\newcommand\bea{\begin{eqnarray}}
\newcommand\eea{\end{eqnarray}}
\newcommand\bseq{\begin{subequations}} 
\newcommand\eseq{\end{subequations}}
\newcommand\bcas{\begin{cases}}
\newcommand\ecas{\end{cases}}
\newcommand{\p}{\partial}
\newcommand{\f}{\frac}

\title{Geometry of the generical cosmological solution before the singularity limit}

\author{Orchidea Maria Lecian}
\email{lecian@icra.it}
\affiliation{DIAEE- Department for Astronautics Engineering, Electrical and Energetics,\\ Sapienza University of Rome, Physics Department, Via Eudossiana, 18- 00184 Rome, Italy.}
\affiliation{Comenius University of Bratislava,\\
Faculty of Mathematics, Physics and Informatics,
Department of Theoretical Physics and Physics Education,\\
Mlynsk\'a Dolina F1, 842 48, Bratislava, Slovakia.}


\begin{abstract}
The generic cosmological solution is analyzed both for the non-asymptotic limit to the cosmological singularity and in the asymptotic limit analytically. The Bianchi I solution and the Bianchi IX solution, described as a sequence of Bianchi I reparameterized solutions, are analyzed with respect to the asymptotic symmetry implied by the space part of the metric tensor. Numerical studies are explained. The semiclassical regime is proposed by using the degrees of freedom for the initial conditions to the Einstein field equations, i.e. those which are not necessarily characterizing for a Bianchi scheme. The appropriate homegeneization techniques and the de-homogenization techniques referred to a generic system of PDE's are discussed and applied to the affine (Misner) space containing the dynamics pertinent to the Hamiltonian problem associated to the solution of the Hamiltonian constraint.\\
The asymptotic limit to the cosmological singularity for the generic cosmological solution is implemented within the asymptotic Kasner parametrization of the BKL approach and for the Misner-Chitre formalism. The EFE in the Misner-Chitre approach imply constraints on the non-asymptotical degrees of freedom, which allow one to define classes of solutions on the anisotropy plane.  
\end{abstract}

\pacs{ 98.80.Jk Mathematical and relativistic aspects of cosmology; 05.45.-a Nonlinear dynamics and chaos; Mathematical methods in physics; Numerical studies of critical behavior, singularities; Partial differential equations; Initial value problem, existence and uniqueness of solutions; General relativity and gravitation}

\maketitle
\section{Introduction\label{section1}}
The generical cosmological solution has been examined to exhibit symmetries in the limit to the singularity. In particular, the asymptotic limit to the cosmological singularity include the neglection of some spacial degrees of freedom; among these, the space gradient of the components of the metric tensor, and the non-diagonal degrees of freedom.\\
\\
Such an analysis describe an oscillating universe in coordinate time .\\
Indeed, in the limit procedures, two different symmetries arise before the limit is implemented, and, also, after the solution to the Hamiltonian constraint, no further physical model and/or confrontation with the analysis of experimental data imply further choices on the description(s).\\
Differently, the quantum implementation of the general cosmological model can imply an avoidance of the singularity without giving up the geometrical spacetime structure and symmetries.\\ 
	 The possibility of the appearance of non- Abelian symmetries as a description for the solution to the Einstein field equations \cite{seri2010}, \cite{OLoughlin:2010sfh} has been exploited for the analysis of the existence of plane-wave solutions for the quantum version of the generic cosmological solution.\\
	The possibility to compare the limit before the singularity to a perturbed FRW model is not included in the confrontation, as the time behavior of the space components of the metric tensor \cite{mccal2003}, \cite{llft} i.e. as well as the derivatives, is different also outside the quantum/experimental uncertainty.\\
	It is possible to analyze the differences of the Misner variables \cite{misner:1969}, and
the simpler realization in the Chitre variables \cite{chitr:72}, chosen for the time evolution with respect to the direct asymptotic Bianchi universes \cite{bel73} with asymptotic Bianchi variables \cite{damour:2002et} after the decomposition of \cite{helgass} in \cite{helga} for the dynamics as far as the consequences on the dynamics due to the quadratic constraint of the Kasner exponents are concerned.\\
	In particular, as expected, such implications do not modify, not only the singularity limit and the corresponding oscillatory regime, but also the overall symmetries of the space geometry \cite{montanibook}. The overall symmetry of the metric tensor is an $SU(2)$ space symmetry; the limit to the singularity allows for the specification of further symmetries, in the simplified models, according to the particular limiting process to be described. Non only the generic cosmological solution admits an $SU(2)$ symmetry as a limit to the singularity, but also a wider class of solutions can exhibit this limit, according to the specification of their properties in time dependence, \cite{Finley:2010hs}, \cite{Babich:1998tz}.\\
\\
In this paper, the limit to the singularity is considered for finding classes of geodesics solutions after the solution to the Hamiltonian constraint, in the time gauge, for the components of the metric tensor, which allow for a parametrization of the initial data of the isotropic degrees of freedom. In particular, the allowed parametrizations for the anisotropy plane are considered, or which the (time-related) expressions for the initial conditions for the parametrization of the anisotropy variables is used to characterize different classes of geodesics.\\
The consequences on the anisotropy plane are analyzed, and the corresponding classes of geodesics are compared, both with the Kasner solution and with the Bianchi IX sequence, also as far as the analysis of the statistics for the initial conditions on the anisotropic degrees of freedom and for isotropy one is concerned \cite{sinai85}, \cite{sinai83}.\\
		
	Analytical solutions for the Hamiltonian constraint are discussed as geodesics for the anisotropy plane. Different classes of analytic solutions are presented, according to the parametrization of the initial data for the isotropic degrees of freedom. The solutions which admit a limit to a Bianchi IX sequence and those who admit a limit to a Kasner geodesic arc of segment are distinguished, according to the non-asymptotic parametrization of the isotropy variable, and to its parametrization in the asymptotic limit to the singularity, as well as for their initial-data problem.\\
		The classes of geodesics can be discussed as the solution for the equations of motion for the Hamiltonian constraint for the analyzed classes of solution to the EFE's. The initial-value problem for the isotropic growth is demonstrated to constitute a variance for the statistical analysis of the corresponding limiting Bianchi IX sequences.\\
		The analytical results for the solutions to the EFE's, the corresponding classes of geodesics in the anisotropy plane and the role of the initial value for the anisotropy degrees of freedom are discussed.\\
		The time evolution of the two different sets of variables, and the quantities chosen for the analysis of the dynamics after the solution to the Hamiltonian constraint are analyzed according to the time evolution, with respect to the parametrized velocities.\\
	The possibility to compare the two different set of variables is allowed as tracked by the role of the initial conditions for the anisotropy variables to allow initial conditions for the isotropic growth of the universe space volume, as numerically calculated. In particular, space of allowed numerical values proves the initial condition for the isotropic growth variable a variance for the comparison on the limit, in which the solutions need correspond.\\
	\\
	The need for analyzing the effects of the initial conditions imposed on the anisotropic degrees of freedom and on the anisotropic ones (on the Misner plane) on the analytical expression of the dynamics for the space of the solutions for the Hamiltonian constraint within arbitrary choices for the Hamiltonian-time variable have already been expressed in the literature \cite{damour:2010sz}, also after the comparison of several numerical investigations \cite{montanirev}. The mathematical tools needed for these analyses are the accomplishment of a dehomogenization\footnote{Here and in the following, the terms 'homogenization' and 'de-homogenization are referred to the analytical techniques (defined on the affine spaces associated to) for the analysis of the system of PDE's and their approximations to a system of ODE's \cite{homog1}, \cite{homog2}, \cite{homog3}; in particular, they do {\slshape{not}} refer to the homogenous cosmological models and the inhomogenous cosmological models named after the Bianchi solutions in the BKL paradigm.} for controlling that the Misner space, which is the space (with arbitrary coordinate choice) of the solutions of the Hamiltonian constraint, in any parametrization of the spherical angular variables, and for a generic (physically-consistent) parametrization of the Hamiltonian time variable, is consistent with inducing Cartesian coordinates on the corresponding Misner planes, on which Cartesian distances can be calculated for the verification of their interpretation with the variance attributed to the Bianchi asymptotic solutions, also in the analytical expressions found during the homogenization analysis and the consequent de-homogenization, as the trajectories on such most general space for he solution of the Hamiltonian constraint is endowed with correct affine (Christoffel) connections.%
	The paper is organized as follows.\\
	In Section (\ref{section2}), the most description of the features of the generical cosmological solution(s) for the EFE's is recalled as far as the non-asymptotical limit is concerned.\\
	In Section (\ref{section3}), the description of the non-asymptotical limit for the Hamiltonian constraint are examined. The Misner-Chitre space, as well as the most general Misner space, are controlled to be accepting de-homogenization for the obtention of a singularity limit within a physical cosmological implementation. The corresponding de-homogenized trajectories are found also without a time gauge.\\
	In Section \ref{section4}, the imposition of the time gauge is controlled to be compatible with the asymptotical symmetries of the Bianchi solutions. The degrees of freedom resulting form the imposition of a time gauge are analyzed to be useful for a statistical characterization of the solutions, as the analytical expressions acquired in the time gauge are compatible to the definition of a variance.\\
	In Section \ref{section5}, the de-homogenized solutions for the EFE's are expressed in the time gauge in the limit to the (cosmological) singularity. The Misner space is analyzed to admit well-posed affine (Christoffel) connections for the solution of the Hamiltonian constraint, which are well behaved both in the cases of the homogenized solution far from the singularity and in the de-homogenized singularity limit; for this, analytical estimates on (geometrical) distances between points lying on geodesics are possible on the (induced Cartesian-coordinatized) Misner plane. The measure of such distances is proven consistent with the analysis of the degrees of freedom present in the time gauge, which stay otherwise not analytically stated. The analytical expression for a variance connecting the de-homogenized trajectories and the Bianchi solutions (which lyi on the same geometrical plane) is demonstrated depending on the initial conditions for the EFE's and on the statistical requirements imposed in the de-homogenization process with respect to the complete Misner plane, where the discrimination for the allocation of the degrees of freedom introduced by the initial conditions either on the (exponentiated expression related to the) coordinate time variable or on the Hamiltonian time variable allows to outline different features for the solutions.\\
	In Section \ref{numericalexamples}, numerical example illustrate the de-homogenization techniques. In particular, the analytical parametrization of the isotropic-growth degrees of freedom obtained for the asymptotic limit in the time gauge of de-homogenized trajectories and their affine connections are compared statistical properties of the Bianchi IX solutions. The application of the requested-order algorithm for the numerical simulations is briefly described.\\
	In Section \ref{section7}, the semiclassical implementation of the model is outlined in Subsection \ref{quantum1} and commented in Subsection \ref{quantum2}: expressions of the de-homogenized solutions are illustrated to be compatible with the expression of semiclassicalization trajectories from a non-Gaussian wavepacket functional postulated after a (possible) quantum phase of the Bianchi Universes, and the comparison with the statistical description arising from a classicalizing quantum regime are conducted.\\
	The comments on the classical implementation of the statistics for the de-homogenized classical Bianchi solutions  introduce the concluding remarks in Section \ref{section8}.	 
\section{Non-asymptotic limit\label{section2}}
It is interesting to analyze the symmetries of the solution of the EFE's in the metric tensor for the generical cosmological solution and for the choices of imposition of initial conditions.\\
This procedure is most generally accomplished not only on the asymptotic-limit analytical expression(s) of the solutions, but, in particular, in the (homogenization) \cite{homog1} Misner space containing the most-generally coordinatized solutions to the original set of the EFE's. 
\subsection{symmetries without the time gauge}
The non-asymptotic limit of of anisotropic oscillating solutions to the Einstein field equations can be inscribed (and analyzed) as an $SU(2)$-invariant (self)dual metric from a classical point of view.\\ 
Such metrics can be expressed in the \cite{Finley:2010hs} line-element form
\begin{equation}\label{timeg}
ds^2=F^2(\vartheta)\omega_1\omega_2\omega_3\left(d\vartheta^2+\sum_{i=1}^{i=3}\frac{(\sum_{r=1}^{r=3}\sigma_i^rdx_r)^2}{\omega_i^2(\vartheta)}\right),
\end{equation}
where the summation over $i$ is performed on the space coordinate, the sum over the index (from the second part of the latin alphabet) $r$ is accomplished on the bein indices. The three functions $\omega$ and the conformal factor $F$ depend on the $0$ coordinate $\vartheta$. This way, the analysis of the generic cosmological solution can be recast by requesting the vanishing of the boundary terms during part integration(s) of the forms $d\tilde{\sigma_i}=\tilde{\sigma}_j\wedge\tilde{\sigma}_k$, with $\tilde{\sigma}_i\equiv\sum_{r=1}^{r=3}\sigma_i^rdx_r$.\\   
They can be most generally stated as containing also an overall conformal factor, depending only on a variable, which, according to an accurate analysis, can be related to the Hamiltonian time variable after the solution of the Hamiltonian constraint.\\
For this analysis, it is necessary to outline that the Lagrangian time variable individuates a direction orthogonal to the hypersurface of the three space degrees of freedom, in the space-time which is solution to the EFE'S. The complete spelling of the indices in (\ref{timeg})ensures that integration by parts of the pertinent Lagrangian density leads to the vanishing of the unwanted terms, as it will be for (\ref{timeg1}) after the consideration of the nextly analyzed relation (\ref{trace}).\\
\subsection{symmetries far from the singularity}
For a Bianchi Universe, the line element is specified as by
\begin{equation}\label{trace}
h_{ij}=R_0^2e^{2\alpha}\sum_{r,s}e^{2\beta^{rs}}\varsigma^{r}_{\ \ i}\varsigma^{s}_{\ \ j}=R_0^2e^{2\alpha}e^{2\beta}_{ij}
\end{equation} 
with $Tr[{\beta}]=0$. The decomposition on the bein $\varsigma$ accomplished with summation on the bein indices $r$, $s$, has been omitted for the symmetries of the Bianchi models. The Kasner solution (corresponding to the Bianchi I solution, which is also employed in the parametrization of the evolution of the other Bianchi solutions) implies also a further quadratic constraint on (suitably normalized) exponents. The isotropic volume element, proportional to the isotropic universe growth 'radius' $R_0$, is not attributed chaos oscillations in vacuum. The presence of a perfect fluid will not be treated here \cite{rugh2}.\\
The case of 'large' volume of the universe \cite{Berger:1989jm}, with respect to the Planck time, describes the cosmological epoch at which a quasi-isotropization mechanism started acting, without any analysis of the isotropic-growth variable.\\
\\
According to this description, the solutions to the EFE's, which admit an $SU(2)$ symmetry in the asymptotic limit to the cosmological singularity therefore can be specified as a sequence of Kasner solutions, for which all the remaining degrees of freedom, i.e. also including parametric expressions and parametrizations of further degrees of freedom, can be expressed, in such a limit, as initial conditions for the isotropy degree of freedom, as it will be quantitatively explicated throughout this work. The initial conditions for the anisotropy degrees of freedom will be demonstrated to undergo only a statistical interpretation, and not a symmetry one, in vacuum, after the analysis of the Misner space.\\
\\
\subsection{Initial conditions and de-homogenization techniques}
The asymptotic behaviour pf the solution to the Einstein field equations if obtained by neglecting the non-diagonal degrees of freedom, both in the case of a homogenous universe, and in the case of inhomogeneities. The presence of microscopic fields \cite{bel73}

 can be compared with the presence of microscopic matter for the initial conditions: such presence can strongly modify, i.e. also by almost cancelling, the oscillatory behaviour(s) and the anisotropic features.\\ 
\\

For the sake of comparison of the non-asymptotic limit without the time gauge and the asymptotic limit within the time gauge, the following normalization for the components of the parametrization of the metric tensor appears appropriate
$e^{p_it}$ $\equiv$ $e^{\frac{\tilde{p}_i-\tilde{p_i}_0}{\tilde{p_i}_0}}e^{\frac{t-\tilde{\Delta t}_0}{\tilde{\Delta t}_0}}$.\\
The diagonal components of the metric tensor, following (\ref{trace}, will be denoted as $\beta^{\mu\mu}$, whose splitting as $\beta_{00}$ and  $\beta_{ii}$, $i=1, 2, 3$ in the asymptotic limit is acquired naturally.\\
 In particular, the set of three values $\tilde{p}_{i,0}$ are the initial conditions for the Einstein field equations, while the coordinate time normalization $\tilde{\Delta t}_0$ is the coordinate time at which the first era ends. Because of the logarithmic change of duration of eras, the latter choice will be effective for the complete duration of the oscillatory regime.\\
\\
The analysis of the initial conditions for a set of ordinary differential equations to which the Einstein field equations- which are a set of partial differential equations for the generic cosmological solution- are reduced in the so-called BKL paradigm \cite{bel73} has been aimed at pointing out the need for a specification of the physical interpretation in the cosmology interpretation the different ways to 're-allocate' the degrees of freedom, which can be considered as redundant for the general cosmological solution.
 The degrees of freedom on the initial conditions which can be considered redundant \cite{sinai85} for the solutions of a system of ordinary ODE find a physical meaning within the suitable de-homogenization techniques \cite{homog1}. For the Bianchi solutions, such de-homogenization techniques correspond to the study of the
parametrization of the Misner-Chitre space. The Misner-Chitre space contains one dimension corresponding to the Hamiltonian time variable in the limit far from the singularity; the de-homogenization of such dimension leads the solution of the Hamiltonian constraint in the time gauge to the asymptotic limit to the cosmological singularity by de-homogenizing the degree of freedom resulting from the initial condition(s) on the isotropic-growth variable \cite{sinai85}. The Bianchi paradigm can be further proved \cite{Finley:2010hs} to admit a precise ($SU(2)$) symmetry in the de-homogenization process, which corresponds to the limit to the (cosmological) singularity.\\
The statistical analysis for the parameters involved in the de-homogenization techniques will be also proven to be compatible with the properties requested for the variance of wavepackets in the Misner-Chitre space, whose semiclassical trajectories have common features with  the properly-de-homogenized Bianchi asymptotical models \cite{homog2}, if non-Gaussian peaked wavepackets are for this purpose built.\\
\\
The need for further analyzing the implication for the trajectories on the Misner plane with (\ref{trace}) receiving a generical parametrization (such as $\Gamma$ in Eq. (\ref{3.3a}) in the Misner (homogenized) space) for the (Hamiltonian) time variable $\alpha$ after the solution of the Hamiltonian constraint was outlined in \cite{damour:2010sz} (as commented there for Eq. (8.10) and for Eq. (8.12)) for the investigation of particular initial conditions on the Misner plane. In the following it will be one of the aims of the analysis to establish the analytical expression or any generic de-homogenized trajectory both in the case of a generic time variable and within the time gauge. The de-homogenization techniques adopted in the following allow to study the well-posed-ness of the analytical expressions of such trajectories, both for the generic Misner space  and for the peculiarities of the Misner-Chitre space, for the , for the projection on the Misner plane, both without the choice of a time gauge and for the time gauge, by the establishment of the requested properties of the affine connections for all these classes of trajectories (solutions to the EFE's) which allow for a suitable matching of the parametrization upon the initial conditions. Furthermore, the explicit expression reconducted to the choice of the generic parametrization of the Hamiltonian time variable will be written both for the appropriate order of approximation and for the desired of the limit to the singularity, and their cosmological implementation will be presented and investigated numerically, for most generic choices of the initial conditions.   
	

 

\section{Detailed analysis of some solutions far from the singularity limit \label{section3}}
The non-asymptotical limit of anisotropic cosmologies towards the cosmological singularity can be discussed according to the $\mu$(-related) variables, of the solution to the Hamiltonian constraint far form the potential.\\
In the following, two different sets of variables for the description of the components of the metric tensor \cite{mccal2003} for the general cosmological solution \cite{llft}  will be investigated with the aim of outlining the role of the initial data problem of the isotropic degree of freedom and of the non-isotropic one as far as the analytical expressions of the geodesics in the anisotropy plane are concerned, according to the EFE's after the solution of the Hamiltonian constraint. In particular, the direct Bianchi I asymptotic decomposition descending from \cite{helga}, as specified in \cite{damour:2002et}, for the Bianchi IX sequences, and the Misner \cite{misner:1969} 
 approach, as well as the Misner-Chitre hypothesis \cite{chitr:72}.\\
\\
The choice of the correct Hamiltonian variables allows one to select the geometrical properties, which have to be investigated \cite{montanibook}.\\
Indeed, two different Hamiltonian sets of variables account for the resolution (description) of the corresponding degree of freedom. Its physical interpretation has therefore to be specified.\\
The two sets of variables differ by the analytic (powerlaw) definition, which implies different (velocity) properties of the solution (with respect to the asymptotics).\\
The different normalization has therefore be interpreted as (different) interference (interaction) properties form the test-mass and the potential wall in the corresponding (quantum gravity) problem.\\
The geometrical placement of the trajectories (from the corresponding initial conditions for the solutions) differs according to the Bianchi models, while their infinite-potential-wall properties (if any) are left unchanged according to the Bianchi classification. For this, the characterization(s) of the Bianchi solution(s) according to Kasner paths is therefore always possible, within the due conditions imposed by both the initial conditions and the specified Bianchi potential.\\
From a non-quantum point of view, the topology of the two spaces being equivalent in three spacetime dimensions, the early universe is described as different as far as the changes in the gravitational interaction are concerned at each Bianchi I sequence.\\

\subsection{Analysis in the Misner plane: the Misner-Chitre decompositions}


The Misner variables and the Misner-Chitre variables offer a parametrization of the components of the metric tensor for the generical cosmological solution, which admit as limits to the cosmological singularity a Kasner description encoded as a Bianchi IX sequence.\\
For comparison with the more recent description \cite{damour:2002et}, the role of the indices $2$ and $3$ have to be exchanged in the following, as already remarked for a comparison with \cite{Chernoff:1983zz} in \cite{damour:2010sz}. For the sake of a comparison of the Cartesian coordinates induced for the goniometric circumference on the Misner plane (as it will be done for Eq.'s (\ref{iwas}) in the following Subsection \ref{subsection3d}), it is relevant to recall that the angular variable defined in \cite{damour:2010sz} differs form the usual choice for the Cartesian plane (it is rotated in the anticounterclock-wise direction of an angle of $\pi/2$).\\
For the Misner variables, the components of the metric tensor obey the EFE condition (\ref{trace}), which here rewrites as
\begin{equation}
tr\frac{d\beta_{ab}}{d\alpha}=0
\end{equation}
and allow for the decomposition
\begin{subequations}\label{qtrace1}
\begin{align}
&q_1=2\alpha+2\beta_{+}+2\sqrt{3}\beta_{-}\\
&q_2=2\alpha+2\beta_{+}-2\sqrt{3}\beta_{-}\\
&q_3=2\alpha-4\beta_{-}
\end{align}
\end{subequations}
which parameterizes the Misner space by the Misner spherical coordinates as
\begin{subequations}\label{trace1}
\begin{align}
&\alpha=-f({\tau})\cosh\zeta\label{3.3a}\\
&\beta_{+}=X=-\frac{1}{2}\frac{d}{d\alpha}\beta_{33}\label{3.3b}\\
&\beta_{-}=Y=-\frac{1}{2\sqrt{3}}\frac{d}{d\alpha}(\beta_{11}-\beta_{22})\label{3.3c}
\end{align}
\end{subequations}
where the polar coordinates  $(X, Y)$, $X\equiv \beta_{+}$, $Y\equiv\beta_{-}$ coordinatize the Misner plane.
In (\ref{trace1}),  
 for $f(\tau)\equiv e^{\tau}$ the original Chitre variables are obtained, and for a generic $f(\tau)=e^{\Gamma(\tau)}$, the Misner-Chitre variables are obtained, which parameterize the corresponding Misner-Chitre space, and define the Misner plane $(X, Y)$ from, after a suitable (more generic) choice of the angular spherical coordinates, i.e. also for $\cosh\zeta=\xi$. Indeed, for any choice of the angular spherical coordinates, the Bianchi dynamics takes place in the unit circle and on the unit circumference on the Misner plane.\\
For the anisotropy degrees of freedom independent of $\alpha$, the following curvilinear abscissa is here rewritten as
\begin{subequations}\label{trace2}
\begin{align}
&\alpha=-\sqrt{\mid \varrho^2+f(\tau)\mid}\\
&\tan\theta=\frac{\beta_{11}-\beta_{22}}{\sqrt{3}\beta_{33}}\label{a}\\
&\varrho=\frac{1}{2\sqrt{3}}\left(\beta_{11}^2+\beta_{22}^2-2\beta_{11}\beta_{22}+3\beta_{33}^2\right)^{1/2}
\end{align}
\end{subequations}
This solution is therefore expressed in the time gauge; the supposed functional independence of $\alpha$ is sufficient for a study of the asymptotic $SU(2)$ symmetry of the model. Nevertheless, such expressions do not consist of a Bianchi IX sequence, as the isotropic degrees of freedom is neglected.\\
To find a Bianchi IX encoded in Kasner solutions it is necessary to let the hypothesis of dependence on $\alpha$ stay, and to impose the constraints on the quantities related to $\alpha$ for the Kasner exponents from the EFE's.\\
\\
\paragraph{preparing for de-homogenizing the Misner space}
The features of the Misner space, which admits generic spherical angular coordinates $\xi$ and generic Hamiltonian time parametrization $\Gamma(\tau)$, are here briefly recalled for the sake of pointing out the relevant degrees of freedom needed for the de-homogenization process \cite{homog1}, also without the imposition of the time gauge.\\
In particular, the solution to the EFE's requires a linear constraint on the derivatives of the $\beta_{ij}$ functions in the exponentiated version of the components of the metric tensor: the (asymptotic) limit to the singularity is expressed therefore by taking into account only the diagonal components, for which the linear Kasner constraint becomes
\begin{equation}\label{traccia}
\sum\frac{d\beta_{ij}}{d\alpha}=0
\end{equation}
in the asymptotic limit.\\
\\
In vacuum, the definition
\begin{equation}\label{kasner2mis}
\sum\left(\frac{d\beta_{ij}}{d\alpha}\right)^2=0
\end{equation}
is not verified identically, but only as a consequence\footnote{It is interesting to remark that the Bianchi I parametrization of the anisotropic degrees of freedom $p_i$ can still be recast in its form in terms of the auxiliary variable as
\begin{subequations}\label{parkas}
\begin{align}
&p_1(u)\equiv-\frac{u}{u^2+u+1},\\
&p_2(u)\equiv\frac{u+1}{u^2+u+1},\\
&p_3(u)\equiv\frac{u(u+1)}{u^2+u+1}.
\end{align}
\end{subequations}
\label{footnote1}} of vanishing right-hand side of the EFE's. This way, the condition (\ref{kasner2mis}) is consistent with a comparison with the role of the quadratic Kasner constraint, because of the role verified (but not necessarily imposed) by the auxiliary variable $u$ (which will later on called also $u^+$) in (\ref{parkas}), as commented in Footnote \ref{footnote1}.\\
\\

No physical information can be grasped, at this stage of the analysis because of its approximation, by considering the exact integration (which will be performed in (\ref{kasner1mis}), (\ref{geod}) and (\ref{ellipses1})).\\
The comparison of the Hamiltonian potential at this stage, and of the numerical simulations of the solutions is nevertheless consistent with the study of the asymptotic limit of the Hamiltonian potential in the associated Hamiltonian problem. In particular,
The Hamiltonian potential for the Bianchi IX solutions is analyzed \cite{montanibook} to reduce to circles for the Bianchi IX model in the anisotropy plane for the Chitre approach, i.e. for $\Gamma(\tau)=\tau$, such that, in the time gauge $\dot{\alpha}=1$, $V(\beta_+, \beta_-)\simeq -3+(\beta_+, \beta_-)^2+\mathcal{O}(\beta^3)$. Within the same assumptions, the Hamiltonian potential is described by the goniometric circle (with respect to the induced Cartesian coordinates) for the Bianchi I solutions.\\
The time gauge is therefore consistent with the hypothesis that the potential be independent of $\tau$ in the limit to the singularity.\\
For a comparison of numerical studies of the evolution of Bianchi IX sequences see, for example, \cite{montanibook} and \cite{montanirev}. 
\subsection{Analysis on the Misner plane: the asymptotic decomposition\label{subsection3d}}
In the most extreme attempt to decompose the metric tensor for the generical cosmological solution for comparison with the Kasner solution is that of the Iwasawa decomposition, for which the Hamiltonian constraint is solved with the purpose to describe a sequence of reparameterized Kasner solutions, matching for the asymptotic limit of the Hamiltonian potential.\\ 
One is allowed to consider the variables $\beta^a=\rho\gamma^a$, normalized as $\beta^\mu=\rho\gamma^\mu$, with $\gamma_\mu\gamma^\mu=1$,and $\rho$ a suitable function, which corresponds to the degrees of freedom, also, before the singularity limit, implied by the quadratic constraints on the Kasner exponents, whose validity proves, on its turn, also before the singularity limit.\\
While a comparison with the parametrization of the Kasner solution is valid, as the generical cosmological solution is explained to consist of a sequence Bianchi I trajectories, a comparison with the parametrization needs further specifications for the time evolution of the metric tensor.\\
Within implementation of the quadratic constraint on the Kasner coefficients in the asymptotic limit, the Misner plane\footnote{The Misner plane is called here the plane coordinatized by the Cartesian coordinates, on which Cartesian distances are calculated.} is coordinatized by circumferences as
\begin{subequations}\label{iwas}
\begin{align} 
&\tan\theta=-\frac{3}{2}\frac{\beta_1-\beta_3}{\beta_3+\beta_1-2\beta_2},\\
&R=\sqrt{6}\left( 1-\frac{3\sum_{i\neq j}\beta_i\beta_j+\sqrt{6}\sum_i\beta_i\frac{3}{2}}{\left(\beta_1+\beta_2+\beta_3+\sqrt{\frac{3}{2}}\right)^2}\right).
\end{align}
\end{subequations}
on which Cartesian coordinates $(X, Y)$ can be induced. The parametrization of geodesics on the Misner plane can be investigated according to the items of information about the initial conditions for the asymptotic limit of the Einstein field equations.\\
The Bianchi I solution admits a parametrization in terms of an auxiliary variable $u$ 
for which the component $u^+$ parameterizes the anisotropic degrees of freedom, while the component $u^-$ the isotropic degrees of freedom due to the initial conditions \cite{sinai85}.\\
\\ 
Given the Kasner parametrization in terms of the accessory $u$ variable, which applies to (\ref{traccia}) and (\ref{kasner2mis}), the dynamics on the Misner plane is limited by the Hamiltonian potentials. In the Bianchi I case, such potential is the cosmological singularity itself, described by the goniometric circle, on the induced Cartesian coordinates (\ref{iwas}). In the Bianchi IX case, as already found in the Misner approach, there coordinatized as $V(\beta_+, \beta_-)$, the Hamiltonian potentials reduce to circles consisting for the induced Cartesian coordinates of the three arcs of circumference centered at $(X-X_{m})^2+(Y-Y_m)^2=R^2_m$, $m=1, 2, 3$, such that $X_{2}=-\sqrt{3}$, $Y_2=1$, $X_{1}=\sqrt{3}$, $Y_1=1$, $X_3=0$, $Y_3=-2$ and radius $R_m\equiv R=\sqrt{3}$ $\forall m$. It is interesting to notice that, within the same geometry, several different realizations are possible \cite{Graham:1990jd} for the description of the Hamiltonian potentials.\\
A de-homogenization on the Hamiltonian potential does not bring any new features, as the corrections to (\ref{iwas1}) (through the solution to the EFE's) are second order, as analyzed for $V(\beta_+, \beta_-)$.\\
\\
The trajectories which describe the asymptotic limit of the evolution of the universe follow the paths of circumferences of radius $R$, with $R\equiv R(u^+,u^-)$  and center $X_c, Y_c$.\\
In particular, the geodesics of the Misner plane further specialize \footnote{ after the relation
\begin{equation}\label{vuuu}
v\equiv \sqrt{\left(u-\frac{u^++u^-}{2}\right)^2-\left(\frac{u^+-u^-}{2}\right)^2}
\end{equation}


as
\begin{subequations}\label{conditioner}
\begin{align}
&X(u,v)=-\frac{1}{2}\sqrt{3}\frac{u^2+v^2+2u}{\left(u+\frac{1}{2}\right)^2+\left(v+\frac{\sqrt{3}}{2}\right)^2},\\
&Y(u,v)=-\frac{1}{2}\sqrt{3}\frac{u^2+v^2-2u-2}{\left(u+\frac{1}{2}\right)^2+\left(v+\frac{\sqrt{3}}{2}\right)^2}.
\end{align}
\end{subequations}
from reversing the relations in \cite{damour:2010sz}. The inverse explicit calculation from the Misner plane to the Lagrangian-density coordinate space was further illustrated also in the first part of \cite{bel2014} resuming previous analyses.} as
\begin{subequations}
\begin{align}\label{iwas1}
&(X-X_c)^2+(Y-Y_c)^2=R^2,\\
&R^2=(X(u_+,0)-X_c)^2+(Y(u_+,0)-Y_c)^2.
\end{align}
\end{subequations}
This, way, the initial conditions for the Einstein field equations are written in an explicit way by setting $u^-=u^-_0$, $v=0$, $u^+$ the value chose for the initial conditions on the isotropic-volume-growth variable.\\  
The coordinates of the centers of the geodesics are parameterized as
 $X_c\equiv X(u_c, r)$, $Y_c\equiv Y(u_c, r)$, where, from \ref{vuuu}, $u_c$ is defined as $u_c\equiv \frac{u^++u^-}{2}$, and $r\equiv \frac{u^+-u^-}{2}$.\\
The expressions (\ref{iwas1}) will be referred to both the geodesics, the description of the Hamiltonian potentials and the de-homogenized expressions for the solutions, as the de-homogenization techniques \cite{homog1} allows to keep the different orders of approximation separated (and separately resummed, when necessary).\\

\subsection{Analysis on the Misner plane: de-homogenized geodesics}
It is now possible to apply the appropriate de-homogenization techniques from the Misner(-Chitre) space to the Misner plane for the integration of the geodesics compatible with the asymptotic limit by analyzing the allowed initial conditions. The role of the isotropic-growth variable will be discussed in further detail.\\
The trace equation Eq. (\ref{traccia}) is valid both in the Misner space and in the Misner-Chitre space.\\
From the analysis of the limit $\alpha\rightarrow\infty$ in the trace equation (\ref{traccia}) for the Misner space and for the Misner-Chitre space, the diagonal components of the metric tensor are integrated as
\begin{equation}\label{kasner1mis}
\lim_{\alpha\rightarrow\infty}\beta_{i}\equiv \alpha p_i +f_i(\alpha),
\end{equation}
with $\beta_{i}\equiv\beta_{jj}$, for $j=1, 2, 3$, and $(f_i)$, $i=1, 2, 3$ generic functions of the isotropic-growth variable obeying the proper limit as in the requested limits for the trace equation, for the linear Kasner constraint and for the quadratic Kasner condition.\\
The functions $f_i(\alpha)$ must therefore obey asymptotic symmetry properties analogous to the Bianchi I parametrization constraints resulting from the EFE's. i.e. that arising from (\ref{kasner1mis}),
\begin{equation}\label{kasner3mis},
\lim_{\alpha\rightarrow\infty}\sum_i f_i(\alpha)=0
\end{equation}
The comparison with the exact asymptotic solution for a generic Bianchi universe is accomplished by the geodesics, which are described as
 \begin{subequations}\label{geod}
\begin{align}
&X(u, u_0, \alpha(0))=f_2(u_0)-\frac{1}{2}\alpha_0 p_2\\
&Y(u, u_0, \alpha(0))=\frac{f_1(u_0)-f_3(u_0)}{3\sqrt{3}}+\alpha_0\frac{p_1-p_3}{3\sqrt{3}}\\
\end{align}
\end{subequations} 
i.e. as ellipses, for which
the initial conditions on the $\alpha$ variable is expressed by the parametrization consequent to the condition (\ref{kasner2mis}) (\ref{trace2}).\\
No physical information is grasped by describing the foci of the ellipses as aligned with a diameter of the corresponding Bianchi I geodesics.\\
\\
A completed analysis of the de-homogenized limit will be obtained in the time gauge, within which the parametrization of the ellipses will be attributed the needed physical information.\\
\paragraph{Comparison of the initial data problems for the geodesics in the anisotropy plane: the isotropic growth initial data}
IWA While the straightforward asymptotic limit of the Bianchi-I-cosmological-model variables obey also a quadratic constraint, which reflects the properties of the Kasner exponents, the Misner variables also obey a quadratic constraint for the derivatives of the components of the metric tensor, which follows form the EFE's.\\
The initial conditions for the solutions to the EFE's in the asymptotic expression for the Bianchi I cosmological solution decomposition are therefore outlined in the expression of the geodesics, where the contribution of the  anisotropic degrees of freedom consists of the $u^+$ value, while the effect of the initial data for the isotropic degrees of freedom is given by the $u^-$ value.\\
The initial value for the anisotropic degree of freedom consists of the numerical value $u_0$, while the initial data for the isotropic degrees of freedom consist of the numerical value of the $\alpha$-variable parametrizations, i.e. $\alpha(u_0)$ and the functions $f_i$ evaluated for $u_0$, related to the $\alpha$ parametrization.\\
For comparison, the initial data for the anisotropic degree of freedom $u_0$ in (\ref{geod}) is the same numerical value of the initial data $u_+$ in (\ref{iwas1}).\\
Differently, the initial data for the isotropic degree of freedom $u_-$ for (\ref{geod}) have the same role of the $\alpha$-related parametrizations $\alpha_0$ and $f_i(0)$ in (\ref{geod}). For the same initial data of the isotropic degree of freedom, the same analytical expression (arcs of circumferences) is found for (\ref{geod}) only in the exact Kasner limit.\\
Accordingly, in the exact Kasner limit, also the same numerical value for the initial data of the isotropic degrees of freedom is re-established.\\
\\
The curvilinear abscissa (\ref{trace2}) admits therefore an exact Kasner limit, but does not consist of a Bianchi IX Kasner-parametrized sequence for the presence of the non-vanishing non-constant $\alpha$ variable.\\
\\
The dynamics on the Misner plane for the asymptotic Bianchi-I-solution variables parametrization is therefore described by a sequence of arcs of circumference, which parameterize after the presence of each potential according to the Kasner parametrization.\\
The dynamics for the Misner-Chitre variables, differently, admits the curvilinear abscissa parametrization (\ref{trace1}).\\
The common $SU(2)$ limit for the dynamics on the Misner plane can be compared by outlining that, for the Misner-Chitre variables, the equations of motions are as well parameterized by the asymptotic degrees of freedom of the metric tensor, consisting of the anisotropic degrees of  freedom plus the initial conditions on the isotropic growth. 

\section{Analysis of the asymptotical limit in the time gauge\label{section4}}
It is possible therefore to analyze the solutions to the EFE's according to the outlined asymptotic $SU(2)$ symmetry, which consists in the possibility to parameterize the solutions according an $SU(2)$-symmetric reparameterized sequence of Kasner solutions. Such parametrization(s) are proved to be useful for the analytical parametrization as a variance (intended for the statistical analysis of the initial conditions, both for the anisotropic degrees of freedom and for the anisotropic ones), whose limiting time-gauge expression is different form the liming one found for the variable $\beta_0$ (and the initial conditions on $\beta_0$).\\
\subsection{Analysis of the symmetries of the time gauge in the asymptotic limit for Bianchi solutions}
 For outlining the features of an $SU(2)$ symmetry, the direction individuated by the time variable is orthogonal to the hypersurface containing the asymptotical spatial degrees of freedom.\\
The $SU(2)$-symmetric asymptotical expressions of the metric tensor (\ref{timeg}) from \cite{Finley:2010hs} allows one \cite{Babich:1998tz} to rewrite the line element s adapted as
\begin{equation}\label{timeg1}
ds^2=F^2(\vartheta)\left(d\vartheta^2+\sum_{i=1}^{i=3}\frac{(\sigma_i^jdx_j)^2}{\Omega_i^2(\vartheta)}\right)
\end{equation}
after a suitable redefinition of the $\vartheta$-dependent functions $\omega_i$ in (\ref{timeg}) for a time gauge expression.\\
\\
The role of the quadratic normalization of the Kasner coefficient is played by the choice of a normalized decomposition for components of the metric tensor as functions of radial and angular variables $\beta^\mu=\rho\gamma^\mu$, with $\gamma_\mu\gamma^\mu=1$.\\
The analytical expression $\lambda=\ln\rho$ for the variables for the solution to the Hamiltonian constraint in \cite{damour:2002et} is consistent with all the requirements of \cite{Babich:1998tz} from \ref{timeg}, but not the most general one fulfilling these requirements. 
\subsection{Isotropy initial conditions as variance\label{sectionisotropy}}
The search for the symmetry properties (as well as for appropriate analytical expressions for the consequent conditions) in the Misner plane after the solution of the Hamiltonian constraint for the Misner variables, as well as for the Misner Chitre ones, can be used  for the selection of initial conditions.\\
Two important cases arise, for which the description of the early universe is different, within the classified $SU(2)$ symmetry properties.\\
\\
Both conditions would imply only a second-order correction to the metric tensor, which do not modify the Ricci scalar.\\
\\ 
It is indeed reasonable to study the implications of the initial conditions after the solution of the Hamiltonian constraint, for the sake of investigating the differences between the Misner variables and their asymptotic-limit expression.\\
In \cite{Furusawa:1985ef}, \cite{Furusawa:1986hf}, the role of the variable $\beta_0$ is commented as the isotropic degree of freedom for the (isotropic) growth of the universe, for which the variance $\Delta_a$ is defined.\\
For this, it is useful to set the difference in the asymptotic limit $\Delta\beta_{ii}\equiv \beta_{ii}-\beta_{ii\ \ 0}$, whose definition applies also to the dimensionally normalized quantities (\ref{trace}).\\
For comparison with (\ref{iwas1}), (\ref{geod}) and (\ref{trace2}), the signature of the metric tensor is the opposite of \cite{Furusawa:1985ef} and \cite{Furusawa:1986hf}.\\
The Misner-Chitre variables (\ref{trace1}), which coordinatize the Misner-Chitre space, are not imposed, by definition, the quadratic Kasner constraint. Their asymptotic limit is nevertheless still consistent with that of variables, on which the quadratic constraint results not as identically verified, but as a consequence  of absence of matter and is imposed on the derivatives of the Misner-Chitre variables.\\
Indeed, the quadratic Kasner constraint ensures an $SU(2)$ asymptotic limit, which is acquired also by the Misner-Chitre variables (\ref{trace1}) in the time gauge, as derived in \cite{Babich:1998tz} for (\ref{timeg1}).\\
For this, it is consistent to impose the quadratic Kasner constraint (\ref{kasner2mis}) 
\begin{equation}\label{kasner5mis} 
tr\left(\frac{d\beta}{d\alpha}\right)=0
\end{equation}
on the derivatives of the set of the de-homogenized geodesics on the Misner plane for 
 obtaining the proper conditions for the de-homogenization of solutions.\\
\\
In the non-asymptotic limit, such a condition is consisting with the definition of the allowed variance $\Delta$ for the sum $\beta_{ii}^2+\beta_{jj}^2$ from Eq. (\ref{kasner5mis}) with respect to the initial conditions, i.e. 
\begin{equation}\label{kasner66mis}
\Delta\beta_{ii}^2+\Delta\beta_{jj}^2\leq0.1
\end{equation}
for which the analytical estimates on the remaining variable $\beta_{kk}$ and for the initial conditions $\beta_{ii\ \ 0}$ is verified in the following subsection (\ref{analyticalestimates}). According to this analysis, the variance $\Delta$  can be applied also to evaluate a variance on the initial conditions with respect to a given step of the evolution, as $\Delta\beta_{ii}^2-\beta_{ii\ \ 0}^2+\Delta\beta_{jj}^2-\beta_{jj\ \ 0}^2\leq0.1$.\\
\\
In the calculation prepared for the analysis of the classical (non-quantum) features of the Bianchi IX solutions in \cite{Furusawa:1985ef}, \cite{Furusawa:1986hf}, the de-homogenization process is not taken onto account, such that the variances calculated are considered only on the asymptotic limit of the solution, i.e. they are considered valid also in the spacetime regions far from the singularity. In particular, the effects of an arbitrary parametrization of the Hamiltonian time variable are not taken into account, differently for the expressions (\ref{geod}) found in the complete Miner space. Indeed, the solutions (\ref{geod}) are more general than those obtained for the semiclassicalization of the trajectories for a Gaussian-peaked wavepacket constructed in the same space.\\
\\
The verification of the quadratic constraint (\ref{kasner5mis}) with the initial conditions on $\beta_0\ \ 0$ leads to
\begin{equation}
\beta_{+}^2+\beta_{-}^2\leq0.5
\end{equation}
as already remarked in the preparation of the numerical results in \cite{Furusawa:1985ef}, \cite{Furusawa:1986hf}. Its numerical value is consistent with those found for (\ref{kasner66mis}).\\
 \\
The variance needed is defined therefore by the coordinate time normalization $\tilde{\Delta t}_0$.\\
The variance discussed in \cite{Furusawa:1985ef}, \cite{Furusawa:1986hf} is therefore that needed to be reconducted the non-asymptotic analysis to a Kasner geodesic arc in an Bianchi IX sequence. No confusion should arise within the variance attributed to the statistical analysis of the probability for the initial condition $u_+$ (i.e. $u_0$ for (\ref{geod})) and $u_-$ in (\ref{iwas1}), and that for the initial condition for the possible analytical expressions of $\alpha$, which will be compared to the statistical analysis of   
\cite{sinai83} after the explicit expression for the de-homogenized solutions.\\
\\
For the sake of illustrating the interpretation of the independence of the statistical probability for Bianchi IX sequences containing a certain number of Bianchi I sequences and the statistical probability for the distribution of the initial conditions for the isotropic-growth variable, numerical examples will be provided in Section \ref{numericalexamples} for non-asymptotic solutions.\\
In particular, here, the variance can be examined and illustrated to depend also on the generical parametrization of the Hamiltonian time variable.\\  
\subsection{variances for asymptotic solutions}
The set of initial conditions prepared to correspond to a variance of $0.1$   to numerically account for the most general initial conditions $\alpha\sim 20$, which are shown to allow for short Bianchi IX sequences, are analyzed in \cite{sinai85} as statistically favored. This is consistent for a de-homogenization admitting $\beta_{ii}^2+\beta_{jj}^2\leq0.1$.
The set of initial conditions with $\alpha\sim4$ can be prepared with the purpose to take numerically into account the most general initial conditions to allow for long Bianchi IX sequences, which are analyzed in \cite{sinai85} as statistically disfavored. This is consistent with a de-homogenization accepting $\Delta\beta_{ii}^2+\Delta\beta_{jj}^2\leq0.01$.\\
\\
The evaluation of the variance for the asymptotic expansion of the de-homogenized solutions (\ref{geod}) in the generically-parameterized space for Hamiltonian solutions is therefore more specific to take into account the role of the initial conditions as split by Theorem (\ref{timeg1}) for the time gauge. The imposition of the initial conditions whose analytical expression has been selected to account for the needed allocation of the degrees of freedom is therefore more general than the asymptotic limit of a solution found considering the simplified Misner Chitre space and, in particular, more effective in discriminating on the effects of such initial condition. Indeed, it allows to distinguish among the cases for which the anisotropy initial conditions are expressed for $\beta_0 \ \ 0$ and for those expressed for any of the $\beta_{ii \ \ 0}$. The statistical analysis applied for the statistically-favored processes according to the number of Bianchi I solutions contained in a Bianchi IX sequence is indeed determined by one of the $\beta_{ii \ \ 0}$ initial conditions, and not on those arising form any degree of freedom imposed on $\beta_0$ (or a related expression $\alpha{\beta_0}$).\\

\section{Analytical expressions of the evolution of the solutions for the asymptotic limit in the time gauge\label{section5}}
It is possible therefore to compare the expression of the $SU(2)$ symmetries of the metric tensor both for the Kasner related classes of solutions (\ref{trace2}) and for the Bianchi IX non-asymptotic analysis of (\ref{geod}) with the exact Kasner limit (\ref{iwas1}) for the relevant ranges of the isotropic degree of freedom.
\subsection{Comparison of the parametrization}
As a further degree of freedom is created by the Hamiltonian formulation, by the choice of the variable, with respect to which the Hamiltonian constraint is problem, which can be chosen as a Hamiltonian time variable (in the Hamiltonian-reduced space, where the time coordinate variable $t$ is not adaptable any more). This is also outlined in the Misner Chitre variables, where such degree of freedom is left to the definition of the function $\Gamma(\tau)$.\\
The comparison of the asymptotic expression of the Bianchi I variables and the Misner variables is preformed on the Misner plane by confronting the role of initial conditions with $\alpha$, and the role of the quadratic constraint on the Kasner coefficient, with respect to the existence of a classification of the metric (\ref{trace}), which allow for the requested $SU(2)$ asymptotic symmetry, without the specification of the general cosmological solution.\\
For this, in 	\cite{Rugh:1994wk}, it has been possible to exclude numerical reconstruction including (isotropic) volume oscillations.\\

\subsection{Analytical solution in the asymptotic limit}
It is possible to compare the two different solutions therefore by imposing a different initial value for the isotropic variable, and to study only the evolution of the anisotropic degrees of freedom, i.e. the Kasner-like solution for the Bianchi-models decomposition, and the equivalent one for the Misner decomposition, within the two proposed ranges of the initial values, which serve as possible values for the variance, by means of which the two decompositions are compared.\\
For this, the corresponding geodesics have been sketched with the same initial conditions for the anisotropic growth.\\
The comparison of the solution to the Hamiltonian constraint for the Misner solution with the asymptotic Kasner parametrization, whose symmetry is admitted by the model (\ref{traccia}), is accomplished by implemented the Kasner solution by the coordinates (\ref{geod}), $(X(u, u_0 \alpha_0), Y(u, u_0, \alpha_0))$ i.e. by considering the trajectories
\begin{equation}
(X(u, u_0, \alpha_0)-X_c)^2+(Y(u, u_0, \alpha_0)-Y_c)^2=R^2
\end{equation} 
where the radius $R$ is that found for the exact $SU(2)$ symmetry. Given the radius as admitted for the initial conditions from reversing the variables (\ref{conditioner}) for (\ref{iwas1}), the degree o freedom $\alpha_0$ accounts for the other intersection point with the potential wall (i.e. with one matched by the direction antiparallel to the direction of motion individuated by the geodesics motion).\\ 
\\
The de-homogenization from the generic Misner space on the Misner plane with respect to an $SU(2)$ symmetry in the limit is therefore parametrized as\\
 \begin{equation}\label{ellipses1}
(a_0+a_1X-X_0)^2+(b_0+b_1Y-Y_0)^2+c_1XY=C^2
\end{equation}
where the degrees of freedom have been split up. The quadratic mixed term $c_1$ is, at this approximation order, accounting for the possible choices of the arbitrary parametrization $\Gamma(\tau)$ in Eq. (\ref{3.3a}).\\
More in particular, the de-homogenization of the Misner space, coordinatized with generical spherical angular coordinates on the solution for the normalized (with respect to the angular velocity at final Hamiltonian time $\alpha$) is therefore as given by the quadratic mixed term $c_1$ in the discriminant and in the Jacobian determinant for quadratic curves (in particular, ellipses). The role of \ref{timeg1} is therefore expressed by the request of a normalization wrt the angular velocity in the limit to the singularity on the Misner plane for an $SU(2)$ symmetry, which induces, on its turn, the determination of the unitary circumference, which coincides with the goniometric circumference after the well-posed-ness of the induction of Cartesian coordinates.\\
The resulting geodesics is therefore an ellipse, which is forced to encode the Kasner parameters by the intersection with the potential lines by construction.\\
\\
It is interesting to summarize \cite{homog1} here that the relation to the initial conditions to the EFE's outlines for the direct limit (without considering the needed homogenization expressions and the de-homogenization techniques) to the singularity (\ref{iwas1}) apply to the de-homogenized limiting solution (\ref{ellipses1}) for the focal distances and for the $C$ parameter (which corresponds to $R$ in (\ref{iwas1})). Any correction to the Hamiltonian potential considered after the de-homogneization is only a second-order correction (with respect to the quadratic-curve parameters $a_0$, $a_1$, $b_0$, $b_1$, $c_1$), as the analytic expression of the Hamiltonian potentials is an exponential of the curvature terms. The de-homogenization technique imposed on the Hamiltonian potential reducing them to the analytical expression of circumferences (i.e. also within the time gauge), as also reviewed in \cite{montanibook}, is conducted therefore by keeping the second order(s) (with respect to the expressions (\ref{iwas1})) separated(ly summed, i.e. or resummed).\\ 
\paragraph{Other analytical solutions for the non-asymptotic limit}
It is possible to compare the two different solutions therefore by imposing a different initial value for the isotropic variable, and to study only the evolution of the anisotropic degrees of freedom, i.e. the Kasner-like solution for the Bianchi-models decomposition, and the equivalent one for the Misner decomposition, within the two proposed ranges of the initial values, which serve as possible values for the variance, by means of which the two decompositions are compared.\\
For this, the corresponding geodesics have been sketched with the same initial conditions for the anisotropic growth.\\
The comparison of the solution to the Hamiltonian constraint for the Misner solution with the asymptotic Kasner parametrization, whose symmetry is admitted by the model (\ref{traccia}), is accomplished by implementing the Kasner solution by the coordinates (\ref{geod}), $(X(u, u_0, \alpha_0), Y(u, u_0, \alpha_0))$ i.e. by considering the trajectories
\begin{equation}
(X(u, u_0, \alpha_0)-X_c)^2+(Y(u, u_0, \alpha_0)-Y_c)^2=R^2
\end{equation} 
where the radius $R$ is that found for the exact $SU(2)$ symmetry. It is consistent with considering also a radius $R$ as evaluated for (\ref{ellipses1})\\
Given the radius $R$ prepared for the de-homogenized asymptotic solution for (\ref{ellipses1}), as a result, the non-asymptotic geodesics rewrites on the Misner plane as
 \begin{equation}\label{ellipsa}
\left(f_2(u_0)-\frac{1}{2}\alpha_0 X-X_c\right)^2+
\left(\frac{f_1(u_0)-f_3(u_0)}{3\sqrt{3}}+\alpha_0\frac{Y}{3\sqrt{3}}-Y_c\right)^2=R^2
\end{equation}
This solution demonstrates that also the non-de-homogenized trajectories projected on the Minser plane, on which Cartesian coordinates can be induced, is endowed with we-posed affine Christoffel connection.
The analytical estimates about distances (evaluated on the Cartesian coordinates induced on the Misner plane) between arcs of geodesics in the previous Subsection \ref{analyticalestimates} remains valid also in the case of circumferences rather then ellipses, because of the $\mathcal{O}$ of such distances imposed by the requested variance (\ref{kasner66mis}), which takes into account also the mixed quadratic term $c_1$ by construction.\\ 
It is always possible to choose the initial condition for $\alpha_0$ such that the analyzed Bianchi statistics is not modified, as it does not influence the (homogenization) parametrization and the (homogenized) parametrization (and the compatibility of the parametrizations) of the auxiliary variable $u^+$ in (\ref{parkas}).\\
The cases in which the statistical analysis of \cite{sinai85} can be retrieved are analyzed as those containing a vanishing quadratic mixed terms $c_1$, after the control that the variances there analyzed are compatible with the estimates of geometrical distances (also, for the sake of direct analysis, expressed in the induced Cartesian coordinates), and after the control that the affine connections also for the non-de-homogenized Misner space parameterized by generic angular spherical coordinates holds.

\subsection{analysis of the initial data: affine connections}
The  $\alpha$ parametrization of the implication of the initial conditions for the isotropic growth on the dynamical system resulting after the solution of the Hamiltonian constraint and the analysis of the consequent $SU(2)$ symmetry requested in the regions of the plane, of the solutions, close to the potential, is intended as \cite{homog1} (de-)homogenization procedures for the initial conditions. Indeed, from the analysis of the affine (Christoffel) connection for the Misner space $(X, Y, \alpha)$, one finds that the geodesics are equivalent only in the exact limit (cfr. with
 \cite{hitchk}, eq. (1a)) $\alpha\equiv0$ .\\
Differently, the auxiliary variable $u$ is not included in the \cite{homog3} (de-)homogenization procedure, as it is defined on a different (sub-)space than the space of the (de-)homogenized solutions \cite{homog1}.\\
For this, it is possible to control that the analytical descriptions of the non-asymptotic limit of the Bianchi I model do not allow the uniqueness in the asymptotic limit from (\ref{timeg}) to (\ref{timeg1}), as the model is singular. Indeed, there is an infinite class of families of parametrizations (\ref{ellipses1}) consistent with the Bianchi initial conditions. Nevertheless, the $SU(2)$ symmetric solution is admitted as a particular case of (\ref{ellipsa}).\\
\\
The difference in the dehomogeneization techniques for a Bianchi I solution and for a trajectory of a Bianchi IX solution are therefore outlined.\\
The initial conditions for the geodesics corresponding to the solutions must be matched at two points, which can be chosen, after the cosmological implementation, at a certain Hamiltonian-variable time $\alpha$, i.e. at the beginning Hamiltonian time $\tau(\alpha)=\tau_0$ and the end of the considered coordinate-time evolution, i.e. the singularity, for which $\alpha$ is not finite, which is determined either by the (asymptotic) limit to the singularity (in the Bianchi I case), or by the presence of an infinite-potential wall (Bianchi IX).\\  
Indeed,
in the case of a Bianchi IX trajectory, one of the initial conditions has to be matched on the potential walls. The further initial condition can be matched either on a (different) potential wall, or on the beginning of the trajectory, defined from the (redundant) isotropic degrees of freedom (corresponding to a particular choice of $\alpha$, i.e. at the beginning $\tau_0$ of the Hamiltonian time) within the potential walls.\\
In the case of a Bianchi I geodesics, it is possible to match both initial conditions on the goniometric circumference, which corresponds to the singularity (limit). The initial conditions for the cosmological implementation of the solution then allows one to match the initial conditions for the trajectories at two points: on the goniometric circumference for the limit to the singularity, and  inside the goniometric circumference at $\alpha=0$.\\ 
\\
It is interesting to understand about the role of the auxiliary variable $u$ (and, in particular, for the anisotropy initial conditions for $u^+$) for the de-homogenization with respect to $\alpha$ in the utilization of the Einstein field equations for a cosmological solution. Indeed, the different parametrization 
 \begin{equation}\label{eqdiff}
\left(f_2(u_0)-\frac{1}{2}\alpha_0 X-X_c\right)^2+
\left(\frac{f_1(u_0)-f_3(u_0)}{3\sqrt{3}}+\alpha_0\frac{Y}{3\sqrt{3}}-Y_c\right)^2=R^2
\end{equation}
is a well-posed de-homogenization scheme \cite{homog1} for the cosmological implementation of the Bianchi solutions, but does not contain the necessary degrees of freedom for the initial conditions.\\
\\
The well-posed-ness of the affine connections for the trajectories in the homogenized space, after the de-homogenization, in the regions of the Hamiltonian description far from the singularity and in the limit close to the singularity, ensures the well-behaved-ness of the matching of all the solutions under the chosen initial conditions by the well-behaved-ness of the quadratic mixed term $c_1$ in the requested approximations form the original solution (corresponding to the original system of PDE's in the generical cosmological solution).
\subsection{analytical estimates\label{analyticalestimates}}
Within the pertinent region of the parametrized solution, the maximum (geodesic) distance $\mathcal{D}$ between the circumference and the ellipse, for the admitted solutions, i.e. for close enough curves at two points $X$ and $X+D$, i.e. at two points of Cartesian distance $D$, with $D<$$\mathcal{D}$, is obtained after expansion of the logarithmic differential as at most as a linear function of the geometric distance, $\textit{d}\sim F_2(X+D)-F_1(D)$; is is as well after expansion majorized by the quadratic inverse of the radius $\mathcal{R}$ of the geodesic circumference on which the $\mathcal{D}$ is evaluated, i.e. $\mathcal{D}$$\simeq1/\mathcal{R}^2$.\\
\\
A Bianchi IX sequence is found within the imposition of the Kasner parametrization on th intersection points with the potential lines. The motion in regions {\it far} from the potential line follows therefore the equations of motion implied by the $\alpha$ variable, for which not only the trajectory is different form the Bianchi I one, but also the velocity is.\\
The asymptotic $SU(2)$ symmetry, common with the Bianchi I solution, is found my matching the ellipsis solution at each match with the Bianchi solution with one potential circumference and is implied by the Hamiltonian constraint. The degree of freedom encoded in the parameter $c_1$ in the expression (\ref{ellipses1}) is needed to select the ellipses tangent to the Bianchi I solutions, i.e. with the same angular velocity at the time where the potential circumference is met. Trajectories arriving at the potential circumferences with a different angular velocities are therefore not solution to the Hamiltonian constraint within the same $\alpha$ parametrization of the initial condition for the isotropic volume Bianchi IX asymptotic solution.   
The possible parametrization of the different non-asymptotical trajectories are implied by the non-trivial dependence on the function $\alpha$ of the Misner coordinates.
\subsection{analytical expression for the variance}
The role of the quadratic mixed term $c_1$ in the time gauge is similar to the action acquired by taking into account the initial conditions on $\beta_0$ without the time gauge. Their numerical values cannot therefore be compared, but only their role within the solution to the EFE's.
\\
According to both the analysis in Subsection (\ref{analyticalestimates}) and those in , the value $\alpha\sim20$ is twice the maximum variance for which the Misner variables and the asymptotic Bianchi solutions can be compared under an allowed statistical (i.e. the uncertainty prepared for the numerical simulation) uncertainty in the limit towards the cosmological singularity. The possibility (of twice the variance) $\alpha\sim 2$ is therefore also another possibility, but, in the present contest, proven as not the most general. The consequences of the particular choice $\alpha\sim 2$ can nevertheless be investigated also within the comparison with the presence of the $c_1$ quadratic mixed term.\\
\\
Here, according to the solution (\ref{geod}) of (\ref{kasner3mis}), the variance $\Delta$  can be applied also to evaluate a variance on the initial conditions with respect to a given step of the evolution, say $A$, as $\Delta\beta_{ii}^2-\beta_{ii\ \ A}^2+\Delta\beta_{jj}^2-\beta_{jj\ \ A}^2\leq0.1$.\\
 According to the analysis of (\ref{ellipses1}), the defined variance $\Delta$  can apply also to calculate variance on the initial conditions with respect to the (Hamiltonian-time-parametrization) initial step of the evolution, as $\Delta\beta_{ii}^2-\beta_{ii\ \ 0}^2+\Delta\beta_{jj}^2-\beta_{jj\ \ 0}^2\leq0.1$.\\
\\
The expression of variances for asymptotic solutions hold in the cases in which the de-homogenized solution (\ref{ellipses1}) exhibit features, due to particular numerical choices of the de-homogenization parameters, which are peculiar, i.e. in the cases they are similar to the deformation of circumferences comparable with those induced by the classicalization of Gaussian-like perturbations. Such cases are the cases in which the quadratic mixing term $c_1$ plays a negligible role, as explained in the previous Section. The analyzed negligible role does not nevertheless depend on the smallness of the numerical value of $c_1$, due to its non-linear effect in the quadratic-curve determinants.\\
I is interesting to remark that those analytical expressions for the variance apply also in the case when the effect of $c_1$ is negligible (such as in the classical calculations prepared for the numerical investigation in \cite{Furusawa:1985ef} and \cite{Furusawa:1986hf}), which can be compared to the non-de-homogenized expressions, such as (\ref{eqdiff}), which have been controlled to admit a well-posed affine connection in the non-dehomogenized space.\\
As a consequence, the variance $\Delta$ depends on the quadratic mixed term $c_1$ and on the initial (with respect to a particular Hamiltonian time) conditions imposed for the anisotropic degrees of freedom and for the isotropic ones, both in the asymptotical de-homogenized case (\ref{ellipses1}) and in the intermediate cases, such as (\ref{eqdiff}) coinciding with a non-dehomogenized solution far from the singularity.\\
The statistical analysis in by \cite{sinai85} is not only here achieved also for the singularity limit (\ref{iwas}), for the regions far from the singularity (\ref{trace2}), for the homogenized equations (\ref{geod}), for their de-homogenized expressions (\ref{ellipses1}) through the imposition of the statistical parameters required by \cite{sinai85} in the Cartesian-coordinatized Misner plane variables (\ref{trace1}) through the reversed variables (\ref{conditioner}).\\
The analyis of the variance here also contains a de-homogenization parameter $c_1$ from (\ref{ellipses1}), which is independent of the previous statistical variables and therefore allows to describe more specific classes of initial conditions, with respect to the choice of allocation of all the possible initial conditions, which are not quotiented out of trace after a redefinition of the metric tensor, the consequences of such redefinition being a redefinition of the boundary terms after (by part) integration for the observables quantities as from a system of holomic coordinates for a Schlesinger equation.\\
Furthermore, the term $c_1$ is also relevant in describing any wanted deformations of the rations among the directrices of the ellipse that generalize the circular trajectories.  
\section{Numerical examples\label{numericalexamples}}
\begin{figure*}[htbp]
\begin{center}
\includegraphics[width=0.4\textwidth]{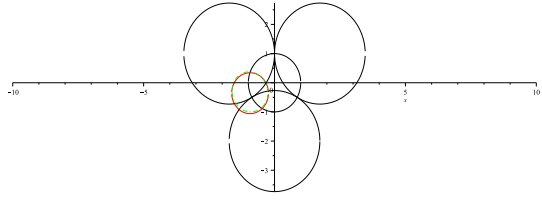}
\caption{\label{figura1} The Misner plane $(X, Y)$ in which the goniometric circumference, corresponding to the cosmological singularity, is depicted. The Hamiltonian problem here associated for a Bianchi I solution is one for which the cosmological singularity corresponds to the goniometric circumference. The Hamiltonian dynamics for the Bianchi IX solution contains three infinite-potential walls, consisting of the three arcs of circumference centered at $(X-X_{m})^2+(Y-Y_m)^2=R^2_m$, $m=1, 2, 3$, such that $X_{2}=-\sqrt{3}$, $Y_2=1$, $X_{1}=\sqrt{3}$, $Y_1=1$, $X_3=0$, $Y_3=-2$ and radius $R_m\equiv R=\sqrt{3}$ $\forall m$.\\
The solution for the asymptotic limit for the corresponding Hamiltonian problem is described by the arcs of circumferences Eq. (\ref{iwas1}), whose de.homogenization, in the (cosmological) singularity limit results as the ellipses Eq. (\ref{ellipses1}).\\
\\
The final trajectory of a Bianchi IX sequence characterized by the initial data on the anisotropic degrees of freedom $u^+=19.7$ and on the isotropic growth degrees of freedom $u^-=-1.3$ corresponds on the Misner plane $(X, Y)$ to the red (gray, solid) circumference of radius $R_{B_{IX}}=0.7011$ and center $X_c=-0.5390$ and     
$Y_c=-0.3583$. A possible dehomogeneized solution, obtained for a statistical variance of $\Delta\beta_{ii}^2+\Delta_\beta{jj}^2\leq0.01$ for which the two asymptotic limits can be compared is depicted from Eq.(\ref{ellipses1}) with the choice of the values $a_0 = 0.001$, $a1 = 0.99$, $b_0 = 0.001$, $b1 = 0.99$, $c1 = 0.001$, $C=0.7011$.}
\end{center}
\end{figure*}

\begin{figure*}[htbp]
\begin{center}
\includegraphics[width=0.4\textwidth]{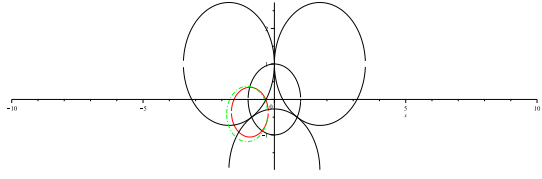}
\caption{\label{figura2}  The last trajectory of a Bianchi IX sequence with the initial data $u^+=19.7$ or the anisotropic degrees of freedom and $u^-=-1.3$ for the isotropic growth degrees of freedom on the Misner plane $(X, Y)$ is the red (gray, solid) circumference of radius $R_{B_{IX}}=0.7011$ and center $X_c=-0.5390$ and     
$Y_c=-0.3583$. A possible dehomogeneized solution, obtained for a statistical variance of $\Delta\beta_{ii}^2+\Delta_\beta{jj}^2\leq0.1$ for which the two asymptotic limits can be compared, is depicted from Eq.(\ref{ellipses1}) with the choice of the values $a_0 = 0.001$, $a_1 = 0.99$, $b_0 = 0.001$, $b_1 = 0.99$, $c_1 = 0.001$, $C=R_{B_{IX}}$.}
\end{center}
\end{figure*}

\begin{figure*}[htbp]
\begin{center}
\includegraphics[width=0.4\textwidth]{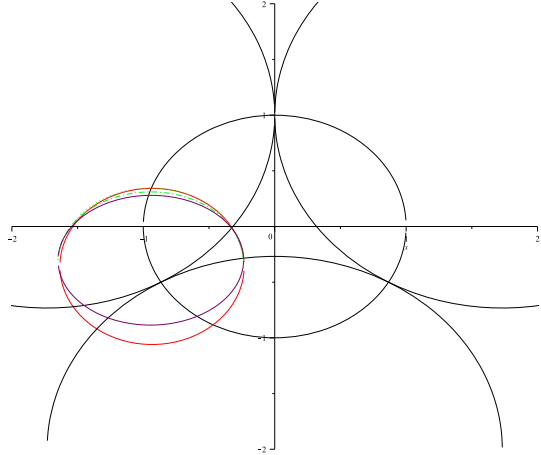}
\caption{\label{figura7} 
The green (dotted) ellipse is obtained for the values $a_0 = 0.001$, $a_1 = 0.99$, $b_0 = 0.001$, $b_1 = 0.99$, $C = R_{B}$, $c_1 = 0.0001$. It corresponds to imposing a variance $0.01$ within the dehomogeneization.\\
The green (dashdotted) trajectory is obtained for the values $a_0 = 0.001$, $a_1 = 0.99$, $b_0 = 0.001$, $b_1 = 1.1$, $c_1 = 0.0001$ from (\ref{ellipses1}), for which  the dehomogeneization techniques in the asymptotical limit process is associated a variance of $\simeq0.1$.\\
The purple (solid, dark gray) ellipses corresponding to the dehomogenization of a circumference trajectory from (\ref{ellipses1}) with $a_0 = 0.001$, $a_1 = 0.99$, $b_0 = 0.001$, $b_1 = 1.2$, $c_1 = 0.01$ allows one to evaluate a variance of $\simeq 0.1$ for the dehomogeneization techniques here adopted.}
\end{center}
\end{figure*}

\begin{figure*}[htbp]
\begin{center}
\includegraphics[width=0.4\textwidth]{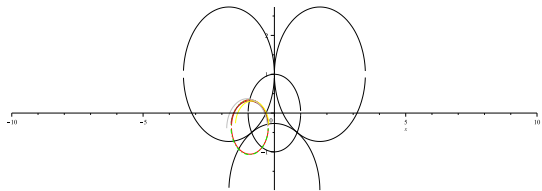}
\caption{\label{figura3}The light-gray solid trajectory is obtained  from Eq. (\ref{ellipses1}) with the values $a_0 = 0$, $a_1 = 0.9$, $b_0 = 0.01$, $b_1 = 0.9$, $c_1 = 0.00001$. The associated variance for the isotropic-growth initial conditions is $\Delta\beta_{ii}^2+\Delta_\beta{jj}^2\leq0.1$ \\
The brown (almost-black, solid) ellipse is obtained from  (\ref{ellipses1}) with  $a_0 = 0.001$, $a_1 = 0.99$, $b_0 = 0.001$, $b_1 = 0.99$, $c_1 = 0.001$. Its dehomogenization corresponds to the asymptotic $SU(2)$ symmetric solution with a variance of $0.01$.\\
The yellow (solid, shading-gray) ellipse is defined as $a_0 = 0.001$, $a_1 = 0.99$, $b_0 = 0.001$, $b_1 = 0.99$, $c_1 = 0.0001$ from the ellipse equation (\ref{ellipses1}). Such choices in the dehomogenization precess allow for a variance $\Delta\beta_{ii}^2+\Delta_\beta{jj}^2\leq0.1$.\\
The different value of $c_1$, corresponding to the mixing term in the quadratic-curve determinants, defines the different ratio among the directrices of the ellipse but do not modify uniquely the variance within the dehomogenization process.\\
The green (dotted) trajectory and the green (dashdotted) trajectory are the same as evaluated in Figure \ref{figura7} for different choises of the jonction of the intersection conditions under the requested variance, for which the points chosen do not necessarily coincide with the goniometric circle, as they are described as part of Bianchi I trajectories.}
\end{center}
\end{figure*}

\begin{figure*}[htbp]
\begin{center}
\includegraphics[width=0.4\textwidth]{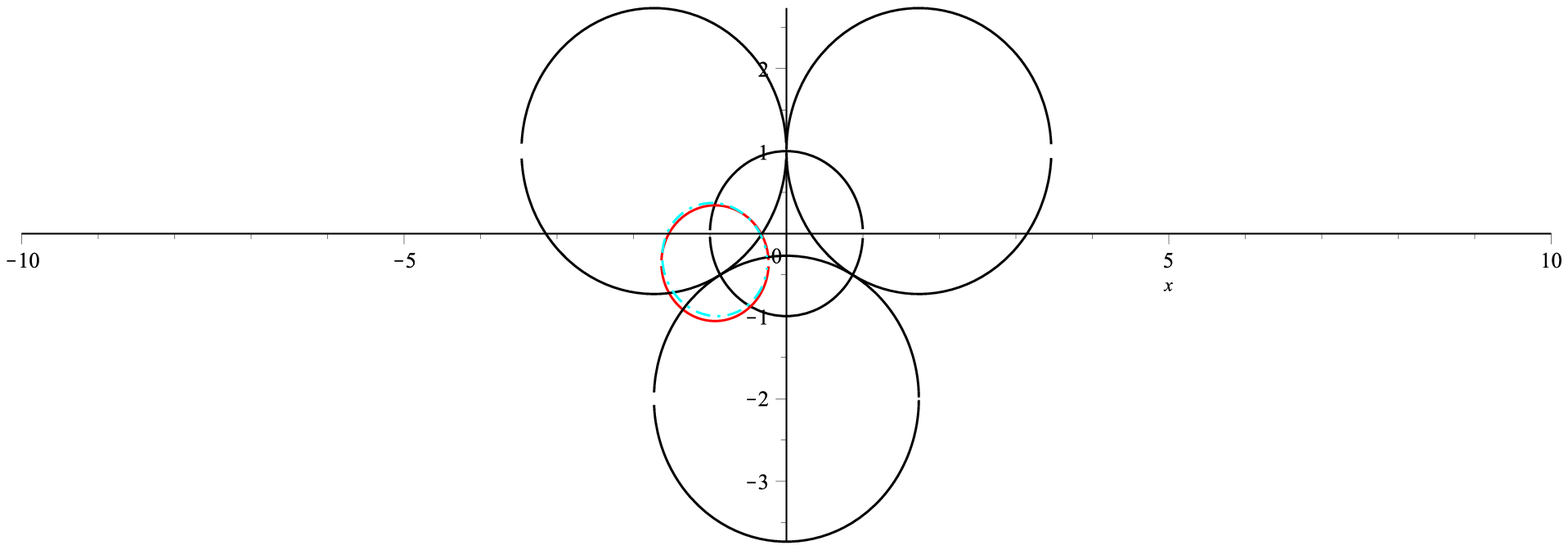}
\caption{\label{figura5} The red (solid) trajectory is the last arc of circumference pertinent to a Bianchi IX sequence with initial conditions $u^+=19.7$ and $u^-=-1.3$ with center $X_c=-0.5390$,    
$Y_c=-0.3583$, and radius $R_{B_{IX}}=0.7011$. The blue (dashdotted) trajectory corresponds to the parameters $a_0 = 0.001$, $a_1 = 0.99$, $b_0 = 0.001$, $b_1 = 0.99$, $C = R_{B_{IX}}$, $c_1 = 0$ in the de-homogenized solution (\ref{ellipses1}).\\
 This case is interesting therefore because, for the vanishing quadratic mixed term, the analysis of the variances applied to the initial conditions for a Bianchi solution in \cite{sinai85} are directly compared with the analytical estimations of (Cartesian) distances of Subsection \ref{analyticalestimates}.\\}
\end{center}
\end{figure*}

It is interesting to apply numerical techniques necessary to produce illustrations of examples for the initial values of the anisotropic degrees of freedom, i.e., as mostly analyzed in the literature, short Bianchi IX sequences and long Bianchi IX sequences.\\
To this aim, one needs cope with the specific algorithms for the solutions of differential equations, i.e., in this case, the forth-order Runge-Kutta method. Regardless to the choice of initial conditions for the anisotropy variables, i.e. indifferently of the choice of cosmological model to be depicted, such as the Kasner solution, short Bianchi IX sequences and long Bianchi IX sequences, the Runge-Kutta method has been here applied by transferring the decision to consider the right number of point in the density (simulation) interval consider for the solution the the differential equation(s) Eq.'s \ref{traccia} for the independent (horizontal abscissa) variable (with respect to which the differential equations are solved) to requesting a solution on an (abscissa) horizontal interval ten times larger than the region where the solution is expected to be found; this makes the numerical simulation gain stability i.e. also in particular with respect to the neighborhood(s) of the regions, in which the Kasner exponents (\ref{parkas}) exhibit a change of parametrization.\\
\\
Numerical examples collect the analytical results.\\
In particular, it is possible to analyze the difference in the requested de-homogeneization techniques for the description of a Bianchi I trajectory and of a Bianchi IX trajectory by exploiting the (redundant) acquired degrees of freedom to match the solutions for the system of partial differential equations which constitute the Einstein field equations at coordinate points, which are subject to a correct cosmological description; more in particular, the degrees of freedom associated to the Hamiltonian time variable have to be chosen as corresponding to the points characterizing the requested Bianchi solution.\\
\\
The resulting de-homogeneization expressions for the Bianchi solutions on the Misner plane have been  analyzed according to the choice of the points where the initial conditions have to be specified for the dehomogeneized solutions (i.e. the ellipses parameterized in (\ref{ellipses1}))
to match the homogenized (asymptotically $SU(2)$-symmetric) Bianchi solution to admit a correct $SU(2)$ asymptotic limit in the cosmological implementation, for which the $SU(2)$ asymptotic limit corresponds to the cosmological singularity.\\
The de-homogeneization techniques here adopted allow one to discriminate between to choice of the initial conditions for the Bianchi IX solution with respect to the Bianchi I solutions as far as the anisotropic degree of freedom and the isotropic-growth degree of freedom are concerned.\\
More in detail, the role of the coordinate time is explained, within the de-homogenization technique, as a de-homogenization of the Hamiltonian time variable $\alpha$ in the asymptotic singularity limit, which allow for the definition of the $SU(2)$ symmetry from Theorem (\ref{timeg1}) in the time gauge.\\
For Bianchi IX sequences, the time end of a trajectory on a potential defined on a geodesics is parameterized by a de-homogenization of the same Hamiltonian time value, at which the dehomogeneized trajectories must match at the same point on the potential wall.\\
The beginning Hamiltonian time $\tau_0$ for a Bianchi IX trajectory is defined by requesting the matching of the de-homogenized trajectories on a (different) potential wall. In the case of the first trajectory of a Bianchi IX sequence, the beginning Hamiltonian time $\tau_0$ in the Misner-Chitre space can correspond as well to a point within the potential domain, but not contained in a potential wall; this feature is thoroughly investigated in \cite{sinai83} and \cite{sinai85} as the role of the auxiliary-parametrization variable (corresponding to) $u^-$ in \cite{bel73}.\\
Different de-homogenization for the Bianchi I solution are matched at the end of the Hamiltonian time $\alpha=\infty$ at the cosmological singularity. The beginning of the Hamiltonian time $\tau_0$ corresponds to a point inside the goniometric circumference (i.e. on the surface delimited by the unit circle) or to a (different) point on the goniometric circumference; this is as well specified by the initial conditions on the isotropic-growth degree of freedom, which is given by the choice of the parameter $u^-$ as in the analysis in \cite{bel73} for \cite{sinai83} and \cite{sinai85}. The difference with respect to the Bianchi IX cosmological solution consists therefore in the (consideration of the) presence of (Hamiltonian) potential walls (in the Misner Chitre space as well) for such a cosmological solution.\\
\\
The numerical examples that follow are aimed at outlining the differences in the dehomogeneization techniques as far as the choices of the corresponding spacetime points on which the conditions have to be matched within a correct $SU(2)$ limit to the (cosmological) singularity and at outlining the role of the choice of the initial conditions on the isotropic-growth degrees of freedom in finding a statistical variance according to which the de-homogenization technique allows for a well-posed $SU(2)$-symmetric limit.\\  
\\
The asymptotic limit of the dynamics on the Misner plane is represented by (segments of) geodesics trajectories. The asymptotic limit for the Hamiltonian description for the Bianchi I solution is delimited by the goniometric circumference accounting for the (cosmological) singularity; the Hamiltonian time at which the dynamics in considered to start on a specific arc of geodesics (containing the point projected on the Misner plane by the value $\alpha(\tau_0)$) can also lye on the unit circle surface (rather than on the goniometric circumference).\\
For the Bianchi I solutions, the initial condition for the anisotropic degree of freedom can be imposed either on the goniometric circumference or for a point in the inside of the corresponding surface. The initial conditions for the isotropic degree of freedom is imposed on the goniometric circumference.\\
In the case of the Bianchi IX solution, the asymptotic limit of the Hamiltonian dynamics is determined by the Hamiltonian potential, consisting of the three arcs of circumference centered at $(X-X_{m})^2+(Y-Y_m)^2=R^2_m$, $m=1, 2, 3$, such that $X_{2}=-\sqrt{3}$, $Y_2=1$, $X_{1}=\sqrt{3}$, $Y_1=1$, $X_3=0$, $Y_3=-2$ and radius $R_m\equiv R=\sqrt{3}$ $\forall m$, which intersect the goniometric circumference only in three points, corresponding to the available access to the (cosmological) singularity.\\
For the Bianchi IX solution, the initial conditions are imposed on the goniometric circumference, even though the corresponding Hamiltonian trajectories do not physically intersect it (except for the case of singular initial conditions).\\
\\
The geodesics shown in Fig. \ref{figura1} represents the end of a long Bianchi IX sequence, i.e. a Bianchi IX solution consisting of several Bianchi I solutions on the Misner plane. The red (solid) line is a circle, corresponding to the asymptotic limit of the last epoch of a Bianchi IX sequence composed of 18 epoch, i.e. for an initial value of $19<u^+<20$, where the trajectory considered is the arc of geodesics within the potential walls inside the unit circle; in  particular, the chosen value here are $u^+=19.70...$ and $u^-=-1.30...$. It is a circle of center $X_c, Y_c$, with $X_c=-0.5390$ and     
$Y_c=-0.3583$ and radius $R_{B_{IX}}=0.7011$.\\
 The green (dashed) line is
the corresponding homogenized trajectory (ellipse) from Eq. (\ref{ellipses1}) with a variance (squared of the possible de-homogenization effects or quantum effects ate semiclassical level)  of $0.01$, and the parameter $C$ chosen as the same value as the radius of the considered dehomogeneized Bianchi trajectory $R_{B_{IX}}$, after keeping the numerical values of the parameters $X_0$ and $Y_0$ from (\ref{ellipses1}) the same as those for $X_c$ and $Y_c$ in the Bianchi solution (\ref{iwas1}) on the Misner plane.\\
The goniometric (unit) circle is in scale $1:1$ in the figure.\\
The numerical values from (\ref{ellipses1}) chosen for this purpose are as following: $a_0 = 0.001$, $a1 = 0.99$, $b_0 = 0.001$, $b1 = 0.99$, $c1 = 0.001$, $C=0.7011$.\\
The horizontal abscissa interval needed for the correct functioning of the Runge-Kutta algorithm is $(-10<X10)$.\\
\\
The geodesics depicted in Fig. \ref{figura2} correspond to the same Bianchi IX (red, solid) trajectory considered in the previous case, i.e. Fig. \ref{figura1}. The dehomogeneization here considered is obtained for a variance of $0.1$, and results as the green (dashed) ellipse. The numerical values chosen for the obtention of the result of such a variance from (\ref{ellipses1}) are the following: $a_0 = 0.001$, $a_1 = 0.99$, $b_0 = 0.001$, $b_1 = 1.1$, $c_1 = 0.001$; the focal distance $C$ is here the same as in the previous case, i.e. $C=R_{B_{IX}}=0.7011$.\\
\\
In Fig. (\ref{figura7}) a Bianchi I geodesics is depicted, for which a Bianchi IX sequence containing a small number of trajectories is obtained, i.e. parameterized by the auxiliary variable $1<u^+<2$. In particular, the circumference trajectory illustrated (red, solid gray) corresponds to the choice $u^+=1.9$, with the initial condition $u^-=-1.30$, as for the previous analyses. These values are chosen is a way  such that value of the radius $C$ of the Bianchi I geodesics trajectory $R_{B_{I}}$ is $R_{B_{I}}\equiv C=0.7011$.\\
This way, it is possible to point out in the analysis the effects of the inhomogeneous degrees of freedom within the initial conditions to the Einstein field equations.\\
Such a trajectory is compared to possible dehomogeneized trajectories, which result as ellipses. The limiting outputs for the dehomogeneization are investigated, to outline the effects of modifying the minor-axis directrix distance (only in one direction, i.e. in the $Y$ direction) by changing the values of the terms $b_1$ in (\ref{ellipses1}) and to shift the center of the ellipses (i.e. by modifying the quadratic mixed term from the quadratic-curve Jacobian determinant) for different values of $b_1$. The ratio between the directrices is further modellized also according the quadratic mixed term $c_1$, whose effects will be more specifically investigated in the following for different cases.\\
Different variances are thus obtained. This way to modify the variance was not explicitly discussed in \cite{Furusawa:1985ef}- \cite{Furusawa:1986hf}, even though both cases are under the same asymptotic limit (\ref{timeg1}) from (\ref{timeg}).\\ 
The green (dotted) ellipse is obtained for the values $a_0 = 0.001$, $a_1 = 0.99$, $b_0 = 0.001$, $b_1 = 0.99$, $C = R_{B}$, $c_1 = 0.0001$. It corresponds to imposing a variance $0.01$ within the dehomogeneization.\\
The green (dashdotted) trajectory is obtained for the values $a_0 = 0.001$, $a_1 = 0.99$, $b_0 = 0.001$, $b_1 = 1.1$, $c_1 = 0.0001$. The variance associated to the dehomogeneization techniques in the asymptotical limit process is $\simeq0.1$.\\
The purple (solid, dark gray) dehomogeneized geodesics is given by the choices of the parameters $a_0 = 0.001$, $a_1 = 0.99$, $b_0 = 0.001$, $b_1 = 1.2$, $c_1 = 0.01$. It corresponds to a variance of $\simeq 0.1$ for the dehomogeneization techniques here adopted.\\
\\
A Bianchi I solution is studied in Fig. \ref{figura3}: in this case, the Bianchi I (Kasner) solution is not picked up from a Bianchi IX sequence, end therefore its parametrization is not requested to match any (boundary) conditions in the asymptotic limit.\\
The Bianchi I solution is one described without the presence of potential walls, for which the boundary conditions have to be matched on the goniometric circle. The boundary conditions correspond, in this cosmological solution, to assign an arbitrary value for the initial conditions for the coordinate variable corresponding to the isotropic degree of freedom.\\
The application of Theorem (\ref{timeg1}) from (\ref{timeg}) allows one to apply de-homogenization techniques to match the de-homogenized solutions in the asymptotic limit to the singularity, i.e. therefore on the goniometric circumference.\\
The de-homogeneization technique here followed for the comparison with a Bianchi IX (part of) trajectory consists in matching the degrees o freedom assigned in the initial conditions on the intersection of the trajectory with the goniometric circumference ($u^+$) as far as the anisotropic degree o freedom in concerned, and on the (maximum) starting point of the trajectory, as far as the isotropic degree of freedom in concerned. The upper bound on the isotropic degree of freedom for the initial conditions is here considered for comparison with the Bianchi I trajectory of a Bianchi IX sequence.\\
In the case of Kasner solution, the initial condition for the beginning Hamiltonian time $\tau_0$ can also coincide to a point on the goniometric circumference $\alpha(\tau_0)\neq\infty$ (where the last part of the inequality corresponds to the Hamiltonian time corresponding to the singularity).\\  
In Fig. (\ref{figura3}), the initial conditions for the Einstein field equations are chosen such as to coincide with the last Bianchi I trajectory of a Bianchi IX sequence, depicted as the red (gray) solid circle; in particular, the initial-value problem has been imposed with the same conditions of the case of Fig (\ref{figura1}), i.e. for $u^+=1.9$. For the peculiarities of the Bianchi I solutions, the choice $u^-=-1.30$ has been followed for the sake of comparison with other cases. In particular, the green (dashdotted) solution is the same corresponding to the variance $0.01$ in Fig. (\ref{figura1}).\\
The light-gray solid trajectory is obtained  form the values $a_0 = 0$, $a_1 = 0.9$, $b_0 = 0.01$, $b_1 = 0.9$, $c_1 = 0.00001$ in (\ref{ellipses1}). Setting $a_0=0$ eliminates the pertinent degree of freedom in the corresponding focal distance of the ellipse and in the constant term. The variance wrt to the circumference solution is of $0.1$.\\
The brown (almost-black, solid) ellipse is defined with  $a_0 = 0.001$, $a_1 = 0.99$, $b_0 = 0.001$, $b_1 = 0.99$, $c_1 = 0.001$ in (\ref{ellipses1}). The variance related to the considered Bianchi I trajectory is $0.01$.\\
The yellow (solid, shading-gray) ellipse, defined as $a_0 = 0.001$, $a_1 = 0.99$, $b_0 = 0.001$, $b_1 = 0.99$, $c_1 = 0.0001$ from (\ref{ellipses1}). The variance associated to the inequality for $\Delta\beta_{ii}^2+\Delta_\beta{jj}^2$ is $0.1$.
 The smaller value of $c_1$, corresponding to the mixing term in the quadratic-curve determinants, defines the different ratio among the directrices of the ellipse (with respect to the previous case, i.e. the brown (almost-black, solid) ellipse differing from the present one for $c_1 = 0.001$).\\
The Bianchi I geodesics matched at the singularity limit but starting at a different Hamiltonian time, whose de-homogenization with respect to the beginning of the Hamiltonian time $\alpha_0\equiv\alpha(\tau_0)$ results as a different initial condition on $f_2(u_0)$ in the de-homogeneization expression (\ref{ellipses1}), and reach the singularity limit with a different angular velocity, the correct $SU(2)$ limit Theorem (\ref{timeg1}) from Theorem (\ref{timeg1}) of the solution being respected. This case is analyzed in the difference between the possibility to set $a_0=0$ (light-gray ellipse) and $a_0\neq0$ (the yellow trajectory and the brown one, that is, the shading-gray trajectory and the almost-black one). The implementation of the corresponding cosmological solutions is therefore that of Bianchi solutions equivalent between the chosen valence of the initial data, i.e. whose (ratios between the) initial conditions on the isotropic-growth degrees of freedom are different for less than $0.1$.\\
In particular, the values for the dehomogeneization on the initial conditions for the yellow (shading-gray) geodesics and for the brown (almost black) geodesics are chosen to match at the asymptotic limit to the (singularity) goniometric circumference, while those for the light-gray trajectory are chosen to match the Bianchi solution on the potential wall at the beginning of the Hamiltonian time for the corresponding Bianchi IX (red, solid) trajectory and for its de-homogenized green (dashdotted) trajectory, where both the green (dotted) trajectory and the green (dashdotted) trajectories are the same as those evaluated for the previous Figure \ref{figura7}.\\
 \\
In Fig. \ref{figura5}
the same case of the geodesics containing the last Bianchi I sequence of a Bianchi IX solution
with initial conditions
$u^+=19.70...$ and $u^-=-1.30...$, of center  $X_c=-0.5390$ and     
$Y_c=-0.3583$ and radius $R_{B_{IX}}=0.7011$ is represented by the red (solid) circumference.\\ 
It is useful to compared it here with the de-homogenized non-asymptotical solution with the quadratic mixed term $c_1$ vanishing. The chosen values for the initial conditions for (\ref{ellipses1}) are $a_0 = 0.001$, $a_1 = 0.99$, $b_0 = 0.001$, $b_1 = 0.99$, $C = R_{B_{IX}}$, $c_1 = 0$; the corresponding de-homogenized geodesics sketched is the blue (dashdotted) geodesics. The relative variance among the two solutions is $0.01$  Such an effect of $c_1=0$ is comparable with the previous cases by a numerical way because of the solution of the quadratic constraint (\ref{kasner5mis}). It has also been confirmed analytically after the results of Section {analyticalestimates}. This case is the most interesting because it allows for a direct comparison of the analytical computation of Cartesian distances between the trajectories an the definition of variances for the anisotropic degree of freedom $u^-$ and their initial conditions analyzed in \cite{sinai85}.\\
\\
The numerical examples here analyzed have outlined most of the details needed in the analysis of the expressions of the geodesics-solution trajectories on the Misner plane from the Misner space, where the latter can coordinatized also with generic angular spherical coordinates, for which the choice of initial conditions, for the anisotropic degrees of freedom as well as for the isotropic ones, is consistent with a statistical analysis, where the variance can be compared with that found for solutions of the same limit of the solutions of the Hamiltonian solution space, whose values can be calculated analytically, due to the validity of analytical estimates for he corresponding Cartesian distances, as due for the well-posed-ness of the Christoffel affine connections, and also numerically. 
\section{Outlook\label{section7}}
It is therefore possible to gain a comparison about the non-asymptotic limit of the two representations.\\
The difference can therefore be described as comparing the two systems as consisting of the same limit considering a prefactor multiplying each variable, which does not modify the both the asymptotic limit and the symmetries of the solutions to the EFE, i.e. the symmetries of the metric tensor.
\subsection{Semiclassicalizing wave-functional in the Misner space\label{quantum1}}
The semiclassical approach to the quantum regime has been described by \cite{Furusawa:1985ef}, \cite{Furusawa:1986hf}, where
the initial wave-packet is hypothesized to be as
\begin{equation}
\psi(\beta_{+}, \beta_{-}, \alpha)=Nexp\left(\vec{p}\dot\vec{\beta}\right)
\end{equation} 
$\vec{\beta}=\beta_{+}, \beta_{-}, \alpha$, i.e. as being suited for being described as a function of the anisotropic variables and of the isotropic growth variable, and $\vec{p}$ the momentum considered for the momentum space of the set of variables $\beta_0$ $\beta_{+}$ and $\beta_{-}$.\\
The quantum regime for the analysis here developed is obtained after the implementation of the Gaussian wavepacket description for the Bianchi limiting expression (\ref{ellipses1}).\\
\\
The analysis of a non-Gaussian wavepacket has been performed in \cite{nongauss1}
and its dispersion relations in \cite{nongauss2}.\\
From these analyses, it is possible to outline that the quadratic-curve mixed term $c_1$ in the de-homogenized expression (\ref{ellipses1}) cannot be re-absorbed within a simple shift of renamed variables, as it would shift the center of the focal distances of the ellipses parameterized in (\ref{ellipses1}).\\
The neglection of the quadratic mixed term has been shown in Fig \ref{figura5}.
The corresponding classical (non-quantum) analysis in the previous Section \ref{numericalexamples} has been aimed at pointing out also numerically that such quadratic term contributes to the calculation of the variance $\Delta_a$ in \cite{Furusawa:1985ef}, \cite{Furusawa:1986hf} regardless to the number of Bianchi I sequences contained in the particular Bianchi IX solution considered.\\
The present implementation of the semiclassical description is relevant in keeping the different regions of the phase space separated.
\\
\subsection{semiclassicalized trajectories on the Misner plane\label{quantum2}}
The set of initial conditions $\Delta_a\leq0.1$ accounts for the most general initial conditions $\alpha\sim 20$, which allow for a small number of Bianchi I sequences, and are proved in \cite{sinai85} as statistically favored.\\
The set of initial conditions $\Delta_a\leq0.01$ are used to analyze a large number of Bianchi I sequences, which are explained as statistically disfavored \cite{sinai85}. Also in this case, the initial conditions allow to statistically compare the two choices of variables, under the proper (twice the) variance, for $\alpha\sim 2$.\\
\\
The variance $\Delta_a$ on which the investigations in \cite{Furusawa:1985ef} and \cite{Furusawa:1986hf} are conducted, is based, as explained at the beginning of the present Section, on different degrees of freedom, i.e. when also $\beta_0$ is considered, and without considering any term having a role similar to that of $c_1$ in (\ref{ellipses1}) among all those that can be imposed on the initial conditions for the EFE's in the asymptotic limit for the Bianchi solution, such as the angular velocity at which the trajectories are described, and on the isotropic-growth variable and on its initial conditions.\\
\\
For this reason, it is useful to use the initial condition for the isotropic growth for the definition of the diagonal independent components of the metric $\beta_{+}$ and $\beta_{-}$, i.e. for three independent degrees of freedom and their initial condition for the EFE's, when the initial conditions do not imply a strong modifications of the Gaussian semiclassical wavepackets, i.e. when the modifications to $\Delta\beta_{ii}+\Delta\beta_{jj}$ are not statistically relevant. More in particular, the consideration applies also when the modifications are not statistically relevant for the chosen values of the parameter quadratic mixed term parameter $c_1$.\\
\\
Within the present analysis, it is possible to outline that the projection on the Misner plane of the trajectory of the semiclassicalized wavepacket corresponds to the simplest (Chitre) choice for the isotropy variable function; accordingly, under the time gauge, the semiclassical wavepacket is expected to be peaked along the classical dehomogeneized trajectory within the dehomogneization-admitted variance imposed by the initial conditions.\\
\\
The consideration of the quadratic mixed term $c_1$ from (\ref{ellipses1}) after the implementation of non-Gaussianly-peaked wavepacket \cite{nongauss1} and \cite{nongauss2} are effective in keeping also the phase space for the dynamics taking place on the Misner space separated.
\section{Concluding Remarks\label{section8}}
The description of the solution to the Hamiltonian constraint corresponds to the comparison of the tow sets of variables at the asymptotic limit.\\
Indeed, the expression of the diagonal degrees of freedom for the metric tensor results in expressing such components of the metric tensor, $g_{ii}$ as function of the auxiliary variable, which parameterizes the solution to the Hamiltonian constrain. Nevertheless, in the Misner case, as well as in the Misner-Chitre case, the quadratic condition on the Kasner coefficients is not exploited.\\
As a result, the further degrees of freedom needed to express the Misner-Chitre parametrization in terms of the Kasner one results in posing such a constraint by allowing for the correspondent degree of freedom. As confirmed after the analytical estimates in Subsection \ref{analyticalestimates}, for the needed interval, the denominator is not divergent, and any series expansion would therefore not be exactly resummed, by keeping the coefficient for each geodesic elements and each Hamiltonian structure separated.\\
\\
\\
The redundant degrees of freedom investigated in the imposition of the trace constraints as the linear one and the verification of the analogous of the quadratic constraint in the non-asymptotic limit coincide with the variance obtained for the corresponding de-homogenization technique.\\
\\
In the present analysis, the dynamics will be investigated, with the components of the metric tensor and the choice of Hamiltonian variables, for which a parametrization of the solutions to the Einstein field equations is explained also by statistical properties.\\
In both cases, the solution of the Hamiltonian constraint leads to the elimination of one degree of freedom, for which the evolution of the dynamics can be described on a two-dimensional surface; as the solution to the Hamiltonian constraint does not admit straight lines on a surface with Cartesian parametrization, it has been pointed out that such a surface is not geometrically Minkowsky flat.\\
\\
It is now possible to compare the two different description within the allowed initial conditions.\\
For this purpose, it is possible to implement such condition(s) by (Taylor) expanding the variable corresponding to such degree of freedom, i.e. within the Misner-Chitre frame, the $e^\tau \sinh\zeta$ factor around the initial value.\\
The asymptotic degrees of freedom of the metric tensor therefore can be expanded by considering the analytical correlations between the request on the initial conditions to the EFE's.\\   
It is possible to chose initial-condition values for the Kasner exponents, for which the importance of the normalization of the Hamiltonian variables becomes outlined.\\
\\
The difference on the non-asymptotic limit to the singularity between the Misner approach and the asymptotic-Bianchi approach is therefore to be looked for in the extra linear extra terms defining the radius, the non-linear terms arising from the different normalization, and the effect of the de-homogenization parameters, also at different steps, on the geodesic motion.\\ 
\\
The comparison within the two description is therefore possible by recovering the differences due to the choice of the time variable with respect to which the Hamiltonian constraint is solved.\\
For this, it is possible to let the solutions to the Einstein field equations in the Misner plane by using the parametrization of the metric exponents according to the Kasner solution, in the (spherical ball neighborhood) vicinity of the singularity, where the two solutions can be compared in the Kasner limit within the $SU(2)$ common symmetry by modifying the Kasner solutions by adding the variance obtained after the solution of the Hamiltonian constraint.\\
\\
The two different evolutions can therefore be considered as statistically equivalently-favored not only at the asymptotic limit: the two sets of variables describe therefore the same physical evolution of the universe not only because they admit the same Kasner parametrization at the singularity limit, but also because they describe the same phenomena within the allowed variance.\\
\\
The comparison of the wavepacket expressed in the asymptotic Bianchi variables with respect to the Misner variables exhibits, for the comparison of the variables, so-called non-Gaussian. The analysis of the expression for the wavepacket cannot be solved by possible renaming the asymptotic-Bianchi-solution variables such that the linear terms are reabsorbed within the renamed variables, as it should imply a shift for the origin of the Misner plane. Differently analyzed, such a shift would imply a change in the initial conditions for the solution to the Einstein field equations, which would imply a different evolution, according to the Kasner parametrization (which would nevertheless stay valid).\\
\\
The present work has been organized as follows.\\
In the Introduction, the motivations of the study have been presented.\\
In Section \ref{section2}, the non-asymptotic limit concerning the three-dimensional space defined by the solution of the Hamiltonian constraint has been investigated. The symmetries which must by obeyed at the asymptotic (with respect to the Hamiltonian time) limit are exposed for the consideration of the Bianchi solutions on the generical cosmological solution, with respect to the analytical expressions of the cosmologically-relevant degrees of freedom within the suitable analytical investigation of the needed de-homogenization techniques.\\
In Section \ref{section3}, the analytical expression of the geodesics for the asymptotical limit on the Misner plane is matched with the solutions in the (homogenized)  Misner-Chitre space and in the de-homogenized Misner plane as far as the initial-value problem is concerned.\\
In Section \ref{section4}, the needed results are put in the time gauge, and the initial-value problem is studied for the choice of the degrees of freedom used in the de-homogenization technique, which allows for the asymptotic limit.\\
In Section \ref{section5}, the results of study the useful degrees of freedom in the de-homogenization are applied to the analytical expressions for the geodesics. Analytical estimates are proposed to account also for the usual assumptions usually presented for the implementation of the quantum regime. 
In Section {\ref{numericalexamples}}, the results are verified also numerically, for which the application of the appropriate-order numerical algorithms is described, and the consequences of the different allowed choices for the initial conditions are illustrated and statistically analyzed.\\
The possibility to construct a semiclassical wavefunctional on which to impose initial conditions according to the dehomogenized solutions is envisaged in , of which the new semiclassical features should arise from non-Gaussian wavepackets are delineated in Section \ref{section7}.\\
Brief concluding remarks end the paper in Section \ref{section8}.

\section*{Acknowledgments} OML is grateful to Prof. Yu. Manin for outlining the importance of reference \cite{Babich:1998tz}. The Author is grateful to Prof. Vladimir Balek for reading the manuscript. This work was partially supported by the National Scholarship Programme of the Slovak Republic- Slovak Academic Information Agency (NS'P- SAIA) (Deadline 30 Apr 2017- Academic Year 2017-2018). Warmest hospitality at Faculty of Mathematics, Physics and Informatics,
Department of Theoretical Physics and Physics Education, Comenius University in Bratislava is heartfully thanked. The software Maple was used for the graphics.
1

\end{document}